\documentclass[twocolumn,amssymb,prb,showpacs,floatfix]{revtex4}
\usepackage{graphicx}
\usepackage{dcolumn}
\usepackage{amsmath}
\usepackage{bm}
\newcommand{\up}{\uparrow}
\newcommand{\down}{\downarrow}

\newcommand{\ea}{\epsilon_{\text{A}}}

\begin{document}

\title{Real-time diagrammatic approach to transport through interacting 
quantum dots with normal and superconducting leads}

\author{Michele Governale$^1$, Marco G. Pala$^2$, and J\"urgen K\"onig$^{1,3}$}
\affiliation{$^1$Institut f\"ur Theoretische Physik III, 
Ruhr-Universit\"at Bochum, 44780 Bochum, Germany\\
$^2$IMEP-LAHC-MINATEC (UMR CNRS/INPG/UJF 5130), 38016 Grenoble, France\\
$^3$Theoretische Physik, Universit\"at Duisburg-Essen, 47048 Duisburg, Germany}

\begin{abstract}
We present a real-time diagrammatic theory for transport through interacting 
quantum dots tunnel coupled to normal and superconducting leads.
Our formulation describes both the equilibrium and non-equilibrium superconducting 
proximity effect in a quantum dot. 
We study a three-terminal transistor geometry, consisting of a single-level quantum dot tunnel coupled to two phase-biased superconducting leads and one voltage-biased normal lead.
We compute  both the Josephson current between the two superconductors and the Andreev current in the normal lead, and analyze their switching on and off as well as transitions between 0- and $\pi$-states as a function of gate and bias voltage.
For the limit of large superconducting gaps in the leads, we describe the formation of Andreev bound states within an exact resummation of all orders in the tunnel coupling to the superconducting leads, and discuss their signature in the non-equilibrium Josephson- and Andreev- current and the quantum-dot charge. 
\end{abstract}



\date{\today}

\maketitle

\section{Introduction}
\label{s1}

Continous advancements in nanofabrication have made it possible to attach superconducting leads to quantum dots. 
The supercurrent through a quantum dot has been measured through dots realized in 
carbon nanotubes \cite{nanotubes} and in InAs nanowires.\cite{nanowires} 
Recently, transport measurements on a single self-assembled InAs quantum dot 
coupled to Al superconducting electrodes have been reported.\cite{buizert07} 

From a theoretical point of view, quantum dots coupled to superconducting 
leads are of great interest, since a very rich physics is expected from the 
combination of superconducting correlations, electron-electron interaction 
and non-equilibrium in the dot.
Subgap transport through a normal-dot-superconductor system is sustained by 
Andreev 
reflection.\cite{fazio98, kang98, schwab99, clerk00, lambert00, cuevas01} 
The Josephson coupling between two superconductors through a quantum dot
has been addressed in the limit of a non-interacting quantum dot
in Ref.~\onlinecite{beenakker92}.
In the opposite limit of a large charging energy, the electrons forming a 
Cooper pair tunnel one by one via virtual dot states,\cite{glazman89,spivak91,rozhkov01} 
which establishes a Josephson current carried by higher-order 
tunneling processes. 
Other aspects of the problem, such as the Kondo regime \cite{glazman89,clerk00bis,avishai03,sellier05,bergeret06,lopez07,karrasch07,nussinov05} or multiple Andreev reflection \cite{cuevas97,johansson99} have also been  addressed. The dependence of the charge in the quantum dot  
on the gate voltage and on the superconducting phase difference has been investigated 
in Ref.~\onlinecite{blatter07}.  
Moreover, numerical approaches based on the non-crossing approximation, \cite{ando95} 
the numerical renormalization group \cite{choi04} and Monte 
Carlo \cite{siano04} have been employed to study transport through this type of systems.
The authors of Ref.~\onlinecite{vecino03} compare different approximation schemes, 
such as mean field and second-order perturbation in the Coulomb interaction. 
The proximity effect in double-dot systems  has been also investigated in different regimes. 
\cite{choi00,bergeret06,lopez07} 
In Ref.~\onlinecite{pala07} the non-equilibrium Josephson and Andreev currents through 
a dot coupled to one normal and two superconducting leads have been studied in the weak-proximity limit, considering only first-order processes in the tunnel coupling with the superconductors. 
In this regime, finite Josephson and Andreev currents can flow only if the dot is driven out of equilibrium. 
The idea of using non-equilibrium to control the behavior of a Josephson 
junction has been proposed\cite{noneqth} and experimentally 
tested\cite{baselmans99} some years ago.

In the present work, we develop a real-time transport theory for an 
interacting quantum dot connected to both superconducting and normal leads.
The theory can be conveniently formulated by means of a diagrammatic language 
and it is suitable for dealing with superconducting correlations, strong 
Coulomb interaction and non-equilibrium due to arbitrary bias voltages
on the same footing.
We demonstrate the use of our formalism for two examples.
First, we study the equilibrium Josephson current between two superconductors 
due to cotunneling through the quantum dot and analyze the formation of a 
$\pi$-state for increasing on-site Coulomb repulsion 
on the dot. 
Second, we consider a transistor geometry with one normal and two
superconducting leads with large superconducting gaps. 
We calculate the non-equilibrium Josephson and Andreev current to all orders 
in the coupling strength with the superconductors, where the quantum dot
is driven out of equilibrium by applying a bias voltage to the normal lead.
This geometry is suitable for performing a spectroscopy of the Andreev bound
states in the interacting quantum dot.

\section{Model and Formalism}
\label{s2}

\subsection{Hamiltonian}

We consider a single-level quantum dot tunnel coupled to both normal and 
superconducting leads. 
The total Hamiltonian of the system is given by
\begin{equation}
  H=H_{\text{D}}+\sum_{\eta}( H_\eta +H_{\text{tunn},\eta}) \, .
\end{equation}
The different (superconducting or normal) leads are labeled by the index 
$\eta$. 
The quantum dot is described by the Hamiltonian of the single-level Anderson 
model,
\begin{equation} 
  H_{\text{D}}=\sum_{\sigma} \epsilon d_{\sigma}^\dagger d_{\sigma}+
  U n_{\uparrow} n_{\downarrow} \, ,
\end{equation}
where $d_\sigma$ ($d_\sigma^\dagger$) 
is the annihilation (creation) operator for an electron in the dot, 
$\epsilon$ denotes the energy of the single-particle level, 
$n_{\sigma}=d_{\sigma}^\dagger d_{\sigma}$ is the number operator for spin 
$\sigma=\uparrow,\downarrow$, and $U$ is the energy cost for double 
occupation.
The leads' electrons are described by the annihilation and creation operators 
$c_{\eta k \sigma}$ and $c_{\eta k \sigma}^\dagger$, respectively.
In addition to the kinetic-energy term 
$\sum_{k \sigma} \epsilon_{k}
c_{\eta k \sigma}^\dagger c_{\eta k \sigma}$ in the Hamiltonian,
there may be a BCS
pair-interaction part $- g_\eta \sum_{k,k'} c_{\eta k \uparrow}^\dagger 
c_{\eta -k \downarrow}^\dagger c_{\eta -k' \downarrow} 
c_{\eta k' \uparrow}$ to account for superconductivity. 
On the one hand, we want to treat the interaction on a mean-field level.
On the other hand, we want to keep track of the total number of electrons,
which will be important for situations with finite bias voltage between
different superconductors.
This can be achieved by representing the lead electrons in terms of
Bogoliubov quasiparticle operators $\gamma_{\eta k\sigma}^{(\dagger)}$
and Cooper-pair annihilation (creation) operators~\cite{Bardeen} 
$S_\eta^{(\dagger)}$ via the Bogoliubov transform
\begin{equation}
  \left( \begin{array}{c} \gamma_{\eta k\uparrow} \\ 
    \gamma_{\eta -k\downarrow}^\dagger \end{array} \right)
  =
  \left( \begin{array}{cc} u_{\eta k} & -v_{\eta k} S_\eta \\ 
    v_{\eta k}^* S^\dagger_\eta & u_{\eta k}^* 
  \end{array} \right)
  \left( \begin{array}{c} c_{\eta k\uparrow} \\ 
    c_{\eta -k\downarrow}^\dagger \end{array} \right).
\end{equation}
with coefficients
\begin{eqnarray}
  u_k &=& \sqrt{ \frac{1}{2} \left( 1 + \frac{\epsilon_k - \mu_\eta}{
    \sqrt{(\epsilon_k-\mu_\eta)^2 + |\Delta_\eta|^2}} \right)}
\\
  v_k &=& e^{i \Phi_\eta} \sqrt{ \frac{1}{2} \left( 1 - 
    \frac{\epsilon_k-\mu_\eta}{\sqrt{(\epsilon_k-\mu_\eta)^2 
	+ |\Delta_\eta|^2}} \right)} 
  \, ,
\end{eqnarray}
where $\mu_\eta$ is the electrochemical potential of lead $\eta$ and
$\Phi_\eta$ the phase of the order parameter 
$\Delta_\eta \equiv g_\eta \sum_k \langle  S^\dagger_\eta
c_{\eta -k \downarrow} c_{\eta k \uparrow} \rangle$.
As a result, the mean-field Hamiltonian for lead $\eta$ reads
\begin{equation}
  H_{\eta}=\sum_{k \sigma} E_{\eta k}
  \gamma_{\eta k \sigma}^\dagger \gamma_{\eta k \sigma} 
  + \mu_\eta N \, ,
\label{hamilton_leads}
\end{equation}
plus an irrelevant constant.
Here, $E_{\eta k} =  \sqrt{(\epsilon_{k}-\mu_\eta)^2 + | \Delta_\eta|^2}$
is the quasiparticle energy, and $N$ is the total number of electrons,
which equals the number of Bogoliubov quasiparticles plus twice the number
of Cooper pairs.
In the case that $\eta$ refers to a normal lead, the order parameter vanishes, 
$\Delta_{\eta}=0$.

The coupling between the dot and the leads is taken into account by the 
tunneling Hamiltonians
\begin{equation}
  H_{\text{tunn},\eta}= 
  V_{\eta} \sum_{k \sigma} \left( c_{\eta k \sigma}^\dagger d_\sigma +
  {\rm H.c.} \right), 
\label{htun}
\end{equation}
where for the sake of simplicity the tunnel matrix 
elements $V_{\eta}$ are considered to be spin and wavevector independent.
The tunnel-coupling strengths are defined as
$\Gamma_{\eta}=2 \pi |V_{\eta}|^2 \sum_k \delta (\omega - \epsilon_k)$, 
which we assume to be energy independent.

\subsection{Diagrammatic Real-Time Technique}
\label{s3}

The main idea of the diagrammatic real-time technique is to integrate out
all the (noninteracting) fermionic degrees of freedom in the leads to arrive 
at an effective description for the reduced system, that is characterized
by the state of the quantum dot and the number of Cooper pairs in the 
superconducting leads.
The Hilbert space of the single-level quantum dot is four dimensional:
the dot can be empty, singly occupied with a spin-up or 
spin-down electron, or doubly occupied.
These are denoted by $|\chi\rangle \in \{ | 0 \rangle, | \uparrow \rangle,
| \downarrow \rangle, | \text{D} \rangle\equiv d^{\dagger}_{\uparrow} 
d^{\dagger}_{\downarrow}|0\rangle \}$, and have energies
$E_0$, $E_\uparrow=E_\downarrow$, and $E_{\text{D}}$, respectively. 
The condensates in the superconducting leads are characterized by the number
of Cooper pairs $|\mathbf{n} \rangle$, relative to some arbitrarily chosen 
reference, where $\mathbf{n}$ is the vector
of Cooper-pair numbers $n_\eta$ for each superconducting lead $\eta$.
The energy contribution from the Cooper-pair condensates is given by
$E_\mathbf{n} = \sum_\eta 2 n_\eta \mu_\eta$.
If all superconducting leads are kept at the same chemical potential then this
energy contribution simply provides a trivial additive constant.
For finite bias voltage between at least two superconducting leads, however,
the total energy depends on how the Cooper pairs are distributed among the
superconducting leads.

We start with the full density matrix of the total system, including the 
quantum dot, the fermionic degrees of freedom of the leads, and the Cooper-pair
condensates.
Since the fermionic degrees of freedom in the leads act as reservoirs, we 
can trace them out to obtain the reduced density matrix $\rho_{\text{red}}$
with matrix elements 
$P_{\xi_2}^{\xi_1} \equiv\langle \xi_1|\rho_{\rm red} |\xi_2\rangle$.

Here, the label $\xi\equiv (\chi,\mathbf{n})$ with energy 
$E_\xi = E_\chi + E_{\mathbf{n}}$ includes both the quantum-dot state $\chi$ 
and the number of Cooper pairs, $\mathbf{n}$, in the leads.
For the diagonal elements of the reduced density matrix we also use the 
notation $P_\xi \equiv P_\xi^\xi$.

\subsubsection{Kinetic equation and current formula}

The dynamics of the reduced density matrix is governed by the kinetic or 
generalized master equation,
\begin{eqnarray}
  \nonumber
  \frac{d}{d t} P^{\xi_1}_{\xi_2}(t) 
  +\frac{i}{\hbar} (E_{\xi_1}-E_{\xi_2}) P^{\xi_1}_{\xi_2}(t)\\
  = \sum_{\xi_1' \xi_2'} \int_{-\infty}^{t} dt' \,
  W^{\xi_1 \xi_1'}_{\xi_2 \xi_2'} (t,t') P^{\xi_1'}_{\xi_2'} (t'), 
  \label{te}  
\end{eqnarray}
where the kernels $W^{\xi_1 \xi_1'}_{\xi_2 \xi_2'} (t,t')$ 
describe transitions due to tunneling.
The current in lead $\eta$ can be written as 
\begin{equation}
  J_{\eta} (t) = -e \sum_{\xi \xi_1' \xi_2'}  
  \int_{-\infty}^{t} dt' \, W_{\xi \xi_2'}^{\xi \xi_1' \eta}(t,t') 
  P_{\xi_2'}^{\xi_1'} (t'),
\label{current3}
\end{equation}
where $W_{\xi \xi_2'}^{\xi \xi_1' \eta} (t,t') \equiv
\sum_s s W_{\xi \xi_2'}^{\xi \xi_1' s \eta}(t,t')$, and 
$W_{\xi \xi_2'}^{\xi \xi_1' s \eta}(t,t')$ is the sum of all kernels that
describe transitions in which in total $s$ electrons are removed from lead 
$\eta$.

Both the generalized master equation and the expression for the current
can be further simplified when all voltages and coupling strengths are kept 
time independent.
The kernels do, then, only depend on the time difference $t-t'$,
and to determine the DC component of the current and all density matrix 
elements we only need the time integrals of the kernels, which we refer to as
generalized rates $W^{\xi_1 \xi_1'}_{\xi_2 \xi_2'}/\hbar
\equiv \int_{-\infty}^t dt' W^{\xi_1 \xi_1'}_{\xi_2 \xi_2'} (t-t')$
and generalized current rates $W^{\xi \xi_1' \eta}_{\xi \xi_2'}/\hbar
\equiv \int_{-\infty}^t dt' W^{\xi \xi_1' \eta}_{\xi \xi_2'} (t-t')$.

The indices $\xi$ contain more information than needed for our purpose.
This is related to the fact that only the change and not the absolute
value of the number of Cooper pairs in each superconducting lead matters,
i.e. the value of the generalized rate $W^{\xi_1 \xi_1'}_{\xi_2 \xi_2'}$ 
does not change when we perform the simultaneous shift 
$\mathbf{n}_{1} \rightarrow \mathbf{n}_{1} + \mathbf{m}$, 
$\mathbf{n}_{2} \rightarrow \mathbf{n}_{2} + \mathbf{m}$, 
$\mathbf{n'}_{1} \rightarrow \mathbf{n'}_{1} + \mathbf{m}$, and
$\mathbf{n'}_{2} \rightarrow \mathbf{n'}_{2} + \mathbf{m}$
for a given vector of additional Cooper-pair numbers $\mathbf{m}$.
After defining 
\begin{eqnarray}
  P_{\chi_2}^{\chi_1} (\mathbf{m}) &\equiv&
  \sum_{ \mathbf{n} }
  P_{(\chi_2,\mathbf{n})}^{(\chi_1,\mathbf{m} + \mathbf{n})}
\\
  W_{\chi_2\chi_2'}^{\chi_1\chi'_1} (\mathbf{m} , \mathbf{m'}) &\equiv&
  \sum_{ \mathbf{n}, \mathbf{n'}}
  W_{(\chi_2,\mathbf{n})(\chi'_2,\mathbf{n'})}
  ^{(\chi_1,\mathbf{m} + \mathbf{n})(\chi'_1,\mathbf{m'} + \mathbf{n'})}
\, ,
\end{eqnarray}
and similarly for the generalized current rates,
we obtain for the stationary current in lead $\eta$
\begin{equation}
  J_{\eta} = -\frac{e}{\hbar} \sum_{\chi \chi_1' \chi_2' \mathbf{n'}}  
  W_{\chi \chi_2'}^{\chi \chi_1' \eta} (\mathbf{0} , \mathbf{n'})
  P_{\chi_2'}^{\chi_1'} (\mathbf{n'})
\, ,
\label{current_n}
\end{equation}
where the matrix elements $P_{\chi_2'}^{\chi_1'} (\mathbf{n'})$ are 
determined from
\begin{equation}
  i \left( E_{\chi_1}-E_{\chi_2} + E_\mathbf{n} 
  \right) P^{\chi_1}_{\chi_2} (\mathbf{n}) =
  \sum_{\chi_1' \chi_2' \mathbf{n'}} 
  W^{\chi_1 \chi_1'}_{\chi_2 \chi_2'} (\mathbf{n},\mathbf{n'})
  P^{\chi_1'}_{\chi_2'} (\mathbf{n'})
\label{master_n}
\end{equation}
together with the normalization condition 
$\sum_\chi P_\chi^\chi(\mathbf{0}) = 1$.
Note that, in Eqs.~(\ref{current_n}) and  (\ref{master_n}), due to conservation of the total number of electrons, only those 
Cooper-pair-number vectors $\mathbf{n}$ appear for which
$\sum_\eta n_\eta$ equals twice the number of dot electrons in state $\chi_2$
minus that in state $\chi_1$ (and the same holds true for $\mathbf{n'}$,
$\chi'_2$, and $\chi'_1$).
The generalized master equations for $P_{\chi_2}^{\chi_1} (\mathbf{n})$ with
all other vectors $\mathbf{n}$, not satisfying the condition stated above, 
decouple and are, therefore, irrelevant. For illustration, let us consider the matrix element 
$P_{0}^{\text{D}}(\mathbf{n})$ in a system with two superconducting leads; for example,  
with $\mathbf{n}=(-1,0)$ or $\mathbf{n}=(-2,1)$ it contributes, while with $\mathbf{n}=(1,0)$ it is irrelevant.

In the special case that all superconducting leads are at the same
chemical potential $\mu_{\text{S}}$, the situation simplifies further.
Due to the fact that in Eq.~(\ref{master_n}) the energy contribution
$E_\mathbf{n}$ is the same for all $\mathbf{n}$ that are compatible with
$\chi_1$ and $\chi_2$, the generalized master equation Eq.~(\ref{master_n})
remains unchanged under the shift 
$\mathbf{n} \rightarrow \mathbf{n} + \mathbf{m}$ and
$\mathbf{n'} \rightarrow \mathbf{n'} + \mathbf{m}$ with
$\sum_\eta m_\eta =0$.
As a consequence, the generalized rates 
$W^{\chi_1 \chi_1'}_{\chi_2 \chi_2'}$ and 
$W_{\chi \chi_2'}^{\chi \chi_1' \eta}$ as well as the solution for 
$P^{\chi_1}_{\chi_2}$ become independent of the Cooper-pair numbers, i.e.,
we can simply drop the arguments $\mathbf{0}$, $\mathbf{n}$ and $\mathbf{n'}$.
It is this limit that we are going to analyze in the results section of this 
paper.

\subsubsection{Time evolution of the reduced density matrix}
\label{s4}
We generalize the real-time diagrammatic approach to transport through 
interacting quantum dots of Ref.~\onlinecite{koenig96} to the case of 
superconducting leads. 
Our goal is to give a diagrammatic prescription to compute the generalized 
rates 
$W^{\chi_1 \chi_1'}_{\chi_2 \chi_2'} (\mathbf{n},\mathbf{n'})$ and
$W_{\chi \chi_2'}^{\chi \chi_1' \eta} (\mathbf{0} , \mathbf{n'})$.
For this, we analyze the time evolution of the reduced density matrix
that we obtain by integrating out the fermionic degrees of freedom in the
leads.

We assume at some initial time $t_0$ (with $t_0 \rightarrow -\infty$) the 
total system to be in a product state of the leads' fermionic degrees of 
freedom (taken at equilibrium) and the degrees of freedom of the reduced 
system.
The time evolution of the reduced density matrix from time $t_0$ to time
$t$ can, then, be described by $P^{\xi_1}_{\xi_2} (t) = 
\sum_{\xi_1' \xi_2'} \Pi^{\xi_1 \xi_1'}_{\xi_2 \xi_2'} (t,t_0) 
P^{\xi_1'}_{\xi_2'} (t_0)$.
The propagator $\Pi^{\xi_1 \xi_1'}_{\xi_2 \xi_2'}(t,t_0)$
can be computed by means of a perturbation expansion in the tunneling 
Hamiltonian $H_{\text{tunn}}=\sum_\eta H_{\text{tunn},\eta}$. 
How this is done has been explained elsewhere\cite{koenig96} 
and here we will limit ourselves to sketch briefly the derivation, thereby 
pointing out the new ingredients due to superconductivity. 
The propagator (starting from and ending at a product state of the leads' 
fermions and the reduced system) is written in interaction representation with 
respect to $H_{\text{tunn}}$ as 
\begin{widetext}
\begin{equation}
  \Pi^{\xi_1 \xi_1'}_{\xi_2 \xi_2'} (t,t') = \text{Tr}_{\text{leads}} 
  \left\{
  \langle\xi_2'(t')|\text{T}_{\text{K}}\left[ |\xi_2 (t) \rangle
    \langle \xi_1 (t)|e^{-\frac{i}{\hbar}\int_{\text{K}_{t'\rightarrow t}} 
      dt'' H_{\text{tunn}} (t'')_{\text{I}}}\right]   
  |\xi_1' (t')\rangle\right\},
\end{equation}
\end{widetext}
being $\text{K}_{t'\rightarrow t}$ the Keldysh contour going from $t'$ to 
$t$ and then backwards to $t'$, $\text{T}_{\text{K}}$ the time-ordering 
operator on the Keldysh contour, $H_{\text{tunn}} (t)_{\text{I}}$ the tunnel 
Hamiltonian in interaction representation, and $\text{Tr}_{\text{leads}}$ 
the trace over the fermionic part of the lead degrees of freedom.

Next, we expand the exponential function in a power series of 
$H_{\text{tunn}}$. 
Finally, we perform the trace over the fermionic lead degrees of freedom by 
means of Wick's theorem. 
This is possible because the Hamiltonians of the leads are quadratic in the 
lead fermionic operators (this applies also to the superconductors in the 
mean-field description adopted here).

For normal leads, only contractions between electron creation and annihilation 
operators are non zero.
They are graphically depicted as tunneling lines (\textit{normal lines}) with 
an arrow going from the vertex $c_{\eta k\sigma}^\dagger d_\sigma$ to the 
vertex $d_\sigma^\dagger c_{\eta k\sigma}$. 
We define the direction of the line such that an electron is removed from the 
dot at the vertex where the line starts and added to the dot at the vertex 
where the line ends.

For the superconducting leads 
two different types of lines appear.
There are again \textit{normal lines}, connecting a vertex
$c_{\eta k\sigma}^\dagger d_\sigma$ with the 
$d_\sigma^\dagger c_{\eta k\sigma}$.
In addition, there are \textit{anomalous lines}, connecting either
$c_{\eta k\sigma}^\dagger d_\sigma$ with
$c_{\eta -k-\sigma}^\dagger d_{-\sigma}$,
or $d^\dagger_\sigma c_{\eta k\sigma}$ 
with $d_{-\sigma}^\dagger c_{\eta -k-\sigma}$.
Due to the convention for the arrow direction introduced above,
the anomalous lines carry two arrows that point towards 
each other (\textit{outgoing anomalous line}) if two annihilation 
operators of dot electrons are involved, and away from each other 
(\textit{incoming anomalous line}) for two creation operators of dot 
electrons. To evaluate the contractions, it is convenient to perform the 
Bogoliubov transform for the lead electron operators. 
Since the Bogoliubov transform involves the operators $S_\eta^{(\dagger)}$,
the number of Cooper pairs in lead $\eta$ may be changed at the tunnel 
vertices.
For each normal and each anomalous superconducting line, there are two possibilities.
The vertices being connected by normal lines either involve no operator
$S_\eta^{(\dagger)}$ or one $S_\eta$ and one $S_\eta^{\dagger}$.
For outgoing (incoming) anomalous lines, either one of the two vertices carries the 
operator $S_\eta^{\dagger}$ ($S_\eta$). 
The normal lines describe quasiparticle tunneling.
Whereas, the anomalous lines describe Andreev tunneling: for an incoming (outgoing) 
anomalous line a Cooper pair breaks (forms) in the lead and its constituents 
enter (leave) the dot at the two vertices.

\begin{figure}
\begin{center}
\includegraphics[width=\columnwidth]{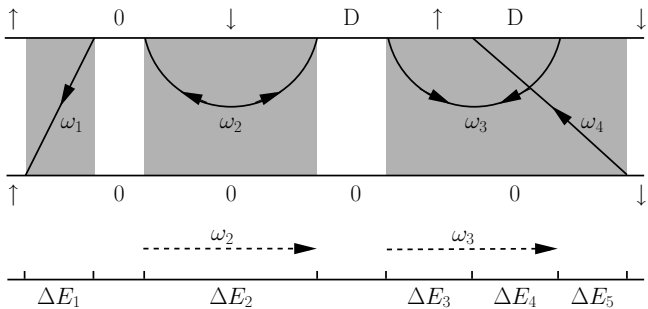}
\caption{
Graphic representation of a contribution to the element $\Pi_{\down \up}^{\down \up}$ of the 
propagator for the reduced density matrix, for the exemplary case of one superconductor with chemical potential 
$\mu_{\text{S}}$. 
The upper and the lower line of the Keldysh contour represents 
the forward and the backward propagation, respectively.
From left to the right we can identify a first-order diagram with a normal line, 
a first-order diagram with an anomalous line, a second-order diagram with a normal and an anomalous lines. Below the diagrammatic representation of the propagator, we show our arbitrary choice for the directions of the anomalous lines and the corresponding energy differences to be used in rule 2, which are given by $\Delta E_1=\omega_1-\epsilon$; $\Delta E_2=-\omega_2-\epsilon+2\mu_{\text{S}}$; $\Delta E_3=-\omega_3-\epsilon+2\mu_{\text{S}}$; $\Delta E_4=\omega_4-\omega_3-2\epsilon-U+2\mu_{\text{S}}$; $\Delta E_5=\omega_4-\epsilon$. }
\label{prop}
\end{center}
\end{figure}

Now we can describe the different contributions to the propagator by means of a graphical representation on the Keldysh contour. An example is shown in Fig.~\ref{prop}, where both normal and 
anomalous lines are present. 

The propagator obeys the Dyson equation
\begin{widetext}
\begin{equation}
  \Pi^{\xi_1 \xi_1'}_{\xi_2 \xi_2'} (t,t') = 
  \Pi^{\xi_1 \xi_1(0)}_{\xi_2 \xi_2} (t,t') \delta_{\xi_1 \xi_1'}  
  \delta_{\xi_2 \xi_2'} +
  \sum_{\xi_1'' \xi_2'' }\int_{t'}^{t} d t''\int_{t''}^{t} dt''' 
  \Pi^{\xi_1 \xi_1(0)}_{\xi_2 \xi_2} (t,t''')
  W^{\xi_1 \xi_1''}_{\xi_2 \xi_2''}(t''',t'')
  \Pi^{\xi_1'' \xi_1'}_{\xi_2'' \xi_2'}(t'',t'), 
\end{equation}
\end{widetext}
where $\Pi^{\xi_1 \xi_1(0)}_{\xi_2 \xi_2}(t,t')$ is the free propagator and we identify  $W^{\xi_1 \xi_1'}_{\xi_2 \xi_2'}(t,t')$ with the irreducible  
part of the propagator, i.e. with the sum of irreducible diagrams going from $t^\prime$ to $t$ %
contributing to the propagator $\Pi^{\xi_1 \xi_1'}_{\xi_2 \xi_2'}(t,t')$, where a diagram is irreducible if any vertical line through the diagrams cuts at least one tunneling line. 
The order in $\Gamma$ of an irreducible diagram is given by the number of tunneling lines present in 
the diagram. Examples of first- and second-order diagrams are shown in Fig.~\ref{prop}.

From now on we concentrate on stationary situations and, hence, we will 
consider the generalized rates. 
These are given by the Laplace transform of 
$W^{\xi_1 \xi_1'}_{\xi_2 \xi_2'} (t-t')$ computed at $z=0^{+}$, i.e 
$W^{\xi_1 \xi_1'}_{\xi_2 \xi_2'}=\hbar\left[\int_{-\infty}^t dt' e^{-z(t-t')} 
W^{\xi_1 \xi_1'}_{\xi_2 \xi_2'} (t-t')\right]_{z=0^{+}}$. 
Furthermore, we only keep the information of Cooper-pair-number differences,
i.e., we formulate the rules for
$W^{\chi_1 \chi_1'}_{\chi_2 \chi_2'} (\mathbf{m} , \mathbf{m'}) 
\equiv \sum_{ \mathbf{n}, \mathbf{n'}}
W_{(\chi_2,\mathbf{n})(\chi'_2,\mathbf{n'})}
^{(\chi_1,\mathbf{m} + \mathbf{n})(\chi'_1,\mathbf{m'} + \mathbf{n'})}$.
The last step does not only reduce the number of matrix elements to be 
considered.
Another virtue is the possibility to combine different contributions.
As mentioned above, for a given superconducting line there are always two 
possibilities to assign operators $S_\eta^{(\dagger)}$ to the two vertices.
Depending on the topology of the reduced diagram, the corresponding 
terms may contribute to {\em different} generalized rates 
$W^{\xi_1 \xi_1'}_{\xi_2 \xi_2'}$ (that differ from each other by the 
number of Cooper pairs) but they always contribute to the {\em same}
$W^{\chi_1 \chi_1'}_{\chi_2 \chi_2'} (\mathbf{m} , \mathbf{m'})$.
It turns out that, since the tunneling strengths $\Gamma_\eta$ are 
independent of energy, the analytic expressions of the two contributions 
always combine nicely, which leads to a rather compact formulation of the 
diagrammatic rules presented below.

As defined above, the arrows of a tunneling line indicate whether an 
electron enters or leaves the dot at a given tunnel vertex.
Furthermore, we want to define an overall direction of each tunneling 
in order to define the sign of the energy carried by the Bogoliubov
quasiparticles.
For normal lines, we will always choose the direction set by the single arrow.
For anomalous lines (that carry two opposite arrows), we pick the direction
arbitrarily, and assign the creation or annihilation of a Cooper pair
to the vertex at which the line direction is opposite to the arrow.

In order to construct a systematic perturbation expansion in the tunnel 
coupling, both the generalized rates and the probabilities are expanded in 
orders of $\Gamma$, i.e 
$W^{\chi_1 \chi_1'}_{\chi_2 \chi_2'} (\mathbf{m} , \mathbf{m'}) = 
W^{\chi_1 \chi_1'(1)}_{\chi_2 \chi_2'} (\mathbf{m} , \mathbf{m'})
+ W^{\chi_1 \chi_1'(2)}_{\chi_2 \chi_2'} (\mathbf{m} , \mathbf{m'})
+ \mathcal{O}(\Gamma^3)$ and 
$P^{\chi_1}_{\chi_2} (\mathbf{m}) =P^{\chi_1(0)}_{\chi_2} (\mathbf{m})
+ P^{\chi_1(1)}_{\chi_2} (\mathbf{m}) + \mathcal{O}(\Gamma^2)$, 
where the superscript indicates the order in $\Gamma$.

\subsubsection{Diagrammatic rules}
\label{ss4}

The rules for evaluating the generalized rates 
$W^{\chi_1 \chi_1'}_{\chi_2 \chi_2'} (\mathbf{m} , \mathbf{m'})$ are:

\begin{enumerate}
\item Draw all topologically different diagrams with fixed ordering of 
the vertices in the real axis. 
The vertices are connected in pairs by tunneling lines carrying energy 
$\omega_i$.
The tunneling lines can be normal or anomalous. 
For each anomalous line choose the direction (forward or backward with respect 
to the Keldysh contour) arbitrarily.

\item  
For each vertical cut between two vertices assign a factor 
$1/(\Delta E +i \eta)$ with $\eta=0^+$, where $\Delta E$
is the difference between the left-going and the right-going energies,
including the energy of the dot states, $E_\chi$, the tunneling lines, 
$\omega_i$, and the energy difference of Cooper-pair condensates, 
$E_{\textbf{n}}$.
The latter is increased (decreased) at each vertex of an outgoing (incoming)
anomalous line at which the arrow is opposite to the arbitrarily chosen line 
direction.

\item 
For each tunneling line assign a factor
$\frac{1}{2 \pi} \Gamma_\eta D_\eta(\omega_i) f^\pm_\eta(\omega_i)$,
where $f^+_\eta(\omega_i) = f_\eta(\omega_i) = [1+\exp(\omega_i-\mu_\eta)/ (k_{{\rm B}} T)]^{-1}$ and $f^-_\eta(\omega_i) = 
1 - f_\eta(\omega_i)$, and
$D_\eta(\omega)=\frac{|\omega-\mu_\eta|}{\sqrt{(\omega-\mu_\eta)^2 -
|\Delta_\eta|^2}} \theta(|\omega-\mu_\eta |-|\Delta_\eta|)$.
The upper (lower) sign applies for lines going backward (forward) with respect 
to the Keldysh contour.
For anomalous lines multiply an additional factor 
$\pm \text{sign}(\omega_i) \frac{|\Delta_\eta|}{|\omega_i|}$.
Moreover, assign a factor $e^{-i \Phi_\eta}$ for an outgoing and
$e^{i \Phi_\eta}$ for an incoming anomalous line.
[For normal leads, only normal lines with $D_\eta(\omega_i) \equiv 1$ appear.]

\item 
Assign an overall prefactor $-i$.\\
Furthermore, assign a factor $-1$ for each\\
a) vertex on the lower propagator;\\
b) crossing of tunneling lines;\\
c) vertex that connects the doubly occupied dot state, 
$| \text{D} \rangle = d^\dagger_{\up} d^\dagger_{\down} | 0 \rangle$, 
to spin up, $| \up \rangle$;\\
d) outgoing (incoming) anomalous tunneling line in which the earlier (later)
tunnel vertex with respect to the Keldysh contour involves a spin up
dot electron.\\ \protect
[The factors in c) and d) arise due to Fermi statistics from the order of 
the dot and lead operators, respectively.]

\item  

For each diagram, integrate over all energies $\omega_i$.
Sum over all diagrams.

\end{enumerate}

The generalized current rates 
$W^{\chi \chi_1' \eta}_{\chi \chi_2'} (\mathbf{0} , \mathbf{m'})$
are evaluated in the following way:

\begin{enumerate}

\item[6.]
Multiply the value of the corresponding generalized rate 
$W^{\chi \chi_1'}_{\chi \chi_2'} (\mathbf{0} , \mathbf{m'})$ with a factor
given by adding up the following numbers for each tunneling 
line that is associated with lead $\eta$:\\
a) for normal lines: $1$ if the line is going from the lower to the 
upper, $-1$ if it is going from the upper to the lower propagator, and $0$
otherwise;\\
b) for anomalous lines: $1$ for incoming lines within the upper and
outgoing lines within the lower propagator, $-1$ for for outgoing lines within 
the upper and incoming lines within the lower propagator, and $0$ otherwise.
\end{enumerate}

The diagrammatic rules are formulated generally enough to account for 
any choice of chemical potentials of the leads, i.e., we allow for any bias
voltages between any pair of leads.
A variety of interesting phenomena, however, shows up already when all 
superconducting leads are kept at the same chemical potential (set to 0 per 
definition), and a nonequilibrium situation is generated only by applying
voltages between the normal leads and the superconductors.
It is this limit that we are going to analyze in the rest of the paper.
In this case, the diagrammatic language simplifies further.
As already indicated above, we simply can drop all information associated with
the Cooper-pair numbers in our diagrammatic rules.

\subsection{Green's functions} 
\label{current_formula}

If all superconducting leads are kept at the same chemical potential,
which we put to $0$ per definition, we can use the simplified diagram, where
the information about the Cooper pairs is ignored.
Such a procedure is identical to having dropped the 
Cooper-pair states from the very beginning in the Hamiltonian, i.e., using the 
Bogoliubov transform without employing the operators $S_\eta^{(\dagger)}$.
For this case, the charge current in lead $\eta$ has been related to the local 
Green's functions of the quantum dot~\cite{clerk00bis,ando95,pala07}
by using the approach of Ref.~\onlinecite{meir-wingreen}.
Here, we report the formula of Ref.~\onlinecite{pala07} which is useful for 
what comes in the following.  
For the sake of keeping the notation compact, we use the  Nambu representation for the dot operators:  $\phi=\left(d_{\up},d^{\dagger}_{\down}\right)^{\text{T}}$. 

The current flowing out of lead $\eta$ is written as the sum of two
contributions, $J_\eta = J_{1\eta} + J_{2\eta}$, with 
\begin{eqnarray}
\label{j1}
  J_{1\eta} &=& \frac{e}{\hbar} \int \frac{d \omega}{2\pi}
  \Gamma_{\eta} D_\eta(\omega)
        {\rm Im} \left\{ {\rm Tr} \left[ \tau_3 
   \left(\mathbf{1} - \frac{\mathbf{\Delta}_\eta}{\omega} \right)
    \right. \right. \nonumber \\ && \left. \left.
    \left(2 \mathbf{G}^{{\rm R}}(\omega) f_\eta(\omega)+ 
    \mathbf{G}^{<}(\omega)\right) \right]\right\} ,\\
  J_{2\eta} &=& -\frac{e}{\hbar} \int \frac{d \omega}{2 \pi} 
  \Gamma_{\eta} \tilde{D}_\eta(\omega)
  {\rm Re}\left\{{\rm Tr}\left[ \tau_3 \frac{\mathbf{\Delta}_\eta}{|\Delta_\eta |}
  \mathbf{G}^{<}(\omega)\right]\right\} ,
  \label{current2}
\end{eqnarray} 
where 
$\mathbf{\Delta}_\eta = \left( \begin{array}{cc} 0 & \Delta_\eta \\ 
\Delta_\eta^* & 0 \end{array} \right)$,
and $f_\eta(\omega)=[1+\exp(\omega-\mu_\eta)/ (k_{{\rm B}} T)]^{-1}$ is the Fermi function, with 
$\mu_\eta$ being the (electro-) chemical potential of lead $\eta$ ($=0$ for the
superconductors), 
$T$ the temperature and $k_{{\rm B}}$ the Boltzmann constant. 
The local dot Green's functions  $\mathbf{G}^{{\rm R}}(\omega)$ and $\mathbf{G}^{<}(\omega)$ 
are matrices in Nambu space, whose components 
$\left( \mathbf{G^{<}}(\omega)\right)_{m, n}$ and $\left( \mathbf{G^{{\rm R}}}(\omega)\right)_{m, n}$ are defined as 
the Fourier transforms of $i\langle\phi^{\dagger}_{n}(0)\phi_m(t)
\rangle$ and $-i \theta(t) \langle \{ \phi_{m}(t),\phi^{\dagger}_{n}(0) \}\rangle$, respectively. 

The two weighting functions $D_\eta (\omega)$ and $\tilde{D}_\eta (\omega)$ are given by 
\begin{eqnarray}
\nonumber
D_\eta(\omega)&=&\frac{|\omega|}{\sqrt{\omega^2 -|\Delta_\eta|^2}}
\theta(|\omega|-|\Delta_\eta|)\\  
\nonumber
\tilde{D}_\eta(\omega)&=&\frac{|\Delta_\eta |}
{\sqrt{|\Delta_\eta|^2-\omega^2}}\theta(|\Delta_\eta|-|\omega|)
\, ,
\end{eqnarray} 
for the superconducting leads, and $D_\eta(\omega) \equiv 1$ and 
$\tilde{D}_\eta(\omega) \equiv 0$ if $\eta$ describes a normal lead.

The current $J_{1\eta}$ involves only excitations energies $\omega$ above the gap.
This is the only contribution in a normal lead, where it reduces 
to the result presented in Ref.~\onlinecite{meir-wingreen}. 
For a superconducting lead, $J_{1\eta}$ has a contribution due to the normal  
elements of the dot Green's function, which describes quasiparticle transport 
and is independent of the superconducting phase difference, 
and a contribution due to the anomalous components of the Green's functions, which is in general phase dependent.   

On the other hand,  $J_{2\eta}$ involves only excitations energies $\omega$ below the gap and 
it describes both Josephson as well as Andreev tunneling.

The above current formula becomes particularly useful in the limit of 
a large superconducting gap ($|\Delta_\eta| \rightarrow \infty$), where 
quasi-particle excitations are inaccessible, $J_{2\eta}$ dominates the 
transport.
In this case, the current in the superconducting lead $\eta$ reads
\begin{equation}
\label{jeta}
  J_\eta = -\frac{2 e}{\hbar} \Gamma_\eta |\langle d_\downarrow d_\uparrow 
  \rangle| \sin (\Psi-\Phi_\eta) \, ,
\end{equation}
where $\langle d_\downarrow d_\uparrow \rangle=|\langle d_\downarrow d_\uparrow \rangle|\exp(i\Psi)$
is the dot pair amplitude.
Equation (\ref{jeta}) has a very simple meaning: it describes the Josephson 
current between the lead with superconducting phase $\Phi_\eta$ and the dot 
with a phase $\Psi$. 
All the complicated physical effects due to the interplay of Coulomb
interaction, coupling to all (normal and superconducting) leads and 
non-equilibrium due to a finite bias voltage between normal and superconducting
leads, are hidden in the dot pair amplitude. 

\section{Results}
 
In the remaining part of the paper, we illustrate our formalism by considering
two examples.

\subsection{Josephson coupling due to cotunneling}
\label{s5}  

First, we analyze the equilibrium Josephson current through a 
superconductor-dot-superconductor system in the limit of weak tunnel coupling.
The lowest-order mechanism that establishes a Josephson coupling between the
superconductors is cotunneling, i.e. the Josephson current starts in second order in the tunnel-coupling strengths $\Gamma_\eta$. 
We consider a symmetric setup with both tunnel-coupling strengths equal to $\Gamma_{\text{S}}$ and $\Delta_{\text{L}}= \Delta_{\text{R}}^*=|\Delta|\exp(i\Phi/2)$.
The two superconductors are kept at the same chemical potential $\mu_{\text{S}}=0$. 
We determine the Josephson current $J_{\text{jos}}=J_{\text{L}}=-J_{\text{R}}$
to second order in 
$\Gamma_{\text{S}}$,
\begin{equation}
  \label{jos2}
  J_{\text{jos}}=
  -\frac{e}{\hbar} \sum_{\chi \chi' \chi''}\left[ 
    W_{\chi \chi'}^{\chi \chi'' \text{L}(2)} 
    P_{\chi'}^{\chi''(0)} + W_{\chi \chi'}^{\chi \chi'' \text{L}(1)} 
    P_{\chi'}^{\chi''(1)}\right]. 
\end{equation}
For the limit $|\Delta|\gg k_{\text{B}} T$ considered here, there is no 
microscopic mechanism in our model to make the dot degrees of freedom relax to equilibrium. 
This situation occurs because the quasiparticles excitation in the 
superconducting leads are not accessible.
In reality, the degrees of freedom of the dot will be coupled to some thermal bath with temperature $T$ and the dot will reach an equilibrium distribution. Hence, we assume that in zeroth order only the diagonal probabilities are non vanishing and they are given by the Boltzmann factors 
\begin{equation} 
  P_\chi^{(0)} = \frac{\exp[-E_\chi/(k_{\text{B}}T)]}{Z} \, ,
\end{equation}
with $E_0=0$, $E_\uparrow = E_\downarrow = \epsilon$, 
$E_{\text{D}} = 2\epsilon+U$ and $Z = \sum_\chi \exp[-E_\chi/(k_{\text{B}}T)]$.
Notice that in the model studied later in Section \ref{deltainf}, the presence 
of a normal lead tunnel coupled to the 
dot provides a   mechanism for the dot to reach equilibrium.    
First, we focus on the regime that both $\epsilon$ and $\epsilon+U$ lie inside the superconducting gap. 
The only non-vanishing first-order correction to the reduced density matrix concerns the off-diagonal element 
$P_{\text{D}}^{0}=\left(P_{0}^{\text{D}}\right)^*$ and it reads
\begin{widetext}
\begin{eqnarray}
\nonumber
P_{\text{D}}^{\text{0}(1)} &=& \frac{i}{2\epsilon+U}\left[W_{\text{D}0}^{00 (1)}P_0^{(0)}+W_{\text{DD}}^{0\text{D} (1)}P_{\text{D}}^{(0)}+2 W_{\text{D}\sigma}^{0\sigma (1)}P_\sigma^{(0)}\right]\\
\label{p0D}
&=& \frac{2}{2 \epsilon+U} \Gamma_{\text{S}} \cos\frac{\Phi}{2} \left\{ 
A\left(\frac{\epsilon}{|\Delta|}\right) P_0^{(0)}-A\left(-\frac{\epsilon+U}{|\Delta|}\right)P_{\text{D}}^{(0)} 
- \left[  A\left(-\frac{\epsilon}{|\Delta|}\right)- A\left(\frac{\epsilon+U}{|\Delta|}\right)\right]P_\sigma^{(0)}\right\},
\end{eqnarray} 
\end{widetext}
where the function $A(z)$ 
is given by
\begin{equation}
A(z)=\frac{1}{\pi} \int_1^{\infty} dx \frac{1}{x+z}\, \frac{1}{\sqrt{x^2-1}}.
\end{equation}

Equation (\ref{p0D}) describes how a finite pair amplitude in the dot can be established in first-order 
in $\Gamma_{\text{S}}$. In fact, $(P_{\text{D}}^{0})^*=P_{0}^{\text{D}}=\langle d_\downarrow d_\uparrow\rangle$ is equal to the pair amplitude in the dot.  

Evaluating the second-order current diagrams $W_{\chi \chi'}^{\chi \chi'' \text{L}(2)}$ (an example is shown in Appendix \ref{appsecond}) and using 
Eq.~(\ref{jos2}), we obtain the following lengthy but complete result for
the Josephson current 
\begin{widetext}
\begin{eqnarray}
\nonumber
J_{\text{jos}} &=&  \frac{2 e}{\hbar} \Gamma_{\text{S}}^2 \sin \Phi \left\{  \left[ \frac{1}{|\Delta|} F\left(\frac{\epsilon}{|\Delta|}\right)
+ \frac{2}{2 \epsilon+U} A^2\left(\frac{\epsilon}{|\Delta|}\right)\right]P_0^{(0)}\right.\\
\nonumber 
& &\left. + 
\left[ \frac{1}{|\Delta|} F\left(-\frac{\epsilon+U}{|\Delta|}\right)
- \frac{2}{2 \epsilon+U} A^2\left(-\frac{\epsilon+U}{|\Delta|}\right)\right] P_{\text{D}}^{(0)}\right.\\ \label{jossecond}
& &\left. -\frac{1}{|\Delta|}\left[ F\left(-\frac{\epsilon}{|\Delta|}\right)+ F\left(\frac{\epsilon+U}{|\Delta|}\right)+4 B\left(\frac{-\epsilon}{|\Delta|},\frac{\epsilon+U}{|\Delta|}\right)  
\right]
P_\sigma^{(0)} \right\},
\end{eqnarray}  
\end{widetext}
where the functions $F(z)$ and $B(z,z^\prime)$ are defined as  
\begin{subequations}
\begin{eqnarray}
\nonumber
F(z)&=&\frac{1}{\pi^2}\int_1^{\infty} dx\, \frac{1}{\sqrt{x^2-1}}\\
& &\int_1^{\infty} dy\, \frac{1}{\sqrt{y^2-1}}\, 
\frac{1}{x+z}\,\frac{1}{x+y}\,\frac{1}{y+z}\\
\nonumber
\hspace{-0.5cm}
B(z,z^\prime)&=&\frac{1}{\pi^2}\int_1^{\infty} dx\, \frac{1}{\sqrt{x^2-1}}\\
& &\int_1^{\infty} dy\, \frac{1}{\sqrt{y^2-1}}\, 
\frac{1}{x+z}\,\frac{1}{x+y}\,\frac{1}{x+z^{\prime}}. 
\end{eqnarray}
\end{subequations}

In the limit $|\Delta| \rightarrow \infty$, all second-order current rates 
vanish and the Josephson current is given by $P_0^{\text{D}(1)}$ multiplied by 
the corresponding first-order current rates, which yields 
\begin{equation}
\label{jos2Dinf}
J_{\text{jos}} =  \frac{e}{\hbar} \Gamma_{\text{S}}^2 \sin \Phi\frac{1}{2\epsilon+U} \left(P_0^{(0)}-P_{\text{D}}^{(0)}\right).  
\end{equation}

The result Eq.~(\ref{jossecond}) for the second-order equilibrium Josephson current is valid 
when both the level $\epsilon$ and $\epsilon+U$ are inside the gap, therefore the limit 
of large interaction $U\rightarrow \infty$ cannot be obtained directly from Eq.~(\ref{jossecond}). 
However, in the limit of large interaction the double occupation of the dot is forbidden, and 
the Josephson current can be obtained by dropping all diagrams involving the 
doubly occupied state 
$|\text{D}\rangle$: 
\begin{eqnarray}
\nonumber
J_{\text{jos}}& =& -\frac{e}{\hbar}\left[W_{00}^{00 \text{L}(2)}P_0^{(0)}+2 W_{\sigma \sigma}^{\sigma \sigma \text{L}(2)}P_\sigma^{(0)} \right]\\
\label{glazmaneq}
&=&\frac{2e}{\hbar}\frac{\Gamma_{\text{S}}^2}{|\Delta|}\sin \Phi \left[ F\left(\frac{\epsilon}{|\Delta|}\right) P_0^{(0)}-F\left(-\frac{\epsilon}{|\Delta|}\right) P_\sigma^{(0)}\right].
\end{eqnarray}
Equation (\ref{glazmaneq}) agrees with the results of Glazman and Matveev.\cite{glazman89}

\begin{figure}
\begin{center}
\includegraphics[width=3.1in]{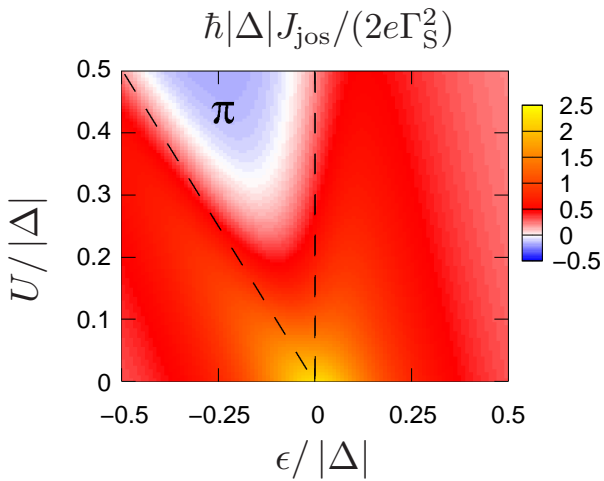}
\caption{(Color online) Density plot of the Josephson current as a function of the level position $\epsilon$ and of the interaction strength $U$. The region where the system behaves as a $\pi$-junction is indicated by 
the symbol $\pi$ in the plot. The dashed lines delimit the region where at zero temperature the dot is singly occupied. 
The other parameters used in the simulation are: $k_{\text{B}} T/|\Delta|=0.05$, 
$\Gamma_{\text{S}}/|\Delta|=0.01$, $\Phi=\pi/2$.}
\label{2ndfig}
\end{center} 
\end{figure}

The Josephson current, Eq.~(\ref{jossecond}), is plotted in Fig.~\ref{2ndfig} 
as a function of gate voltage and interaction strength.
We find, in agreement with Ref.~\onlinecite{nanowires},
the formation of a $\pi$-state for gate voltages such that 
$-U\lesssim\epsilon\lesssim 0$, with the transitions being smeared out by 
temperature.

\subsection{Andreev-level spectroscopy}
\label{deltainf}

\begin{figure}
\begin{center}
\null\vspace{0.5cm}
\includegraphics[width=3.1in]{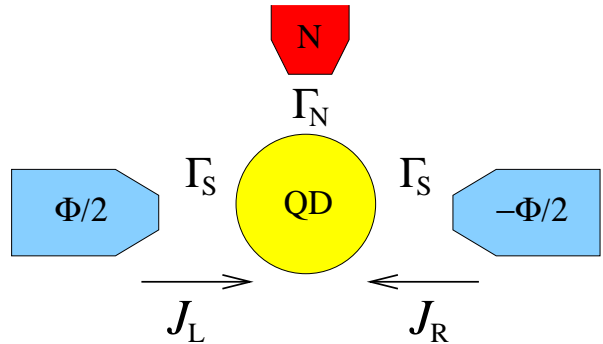}
\caption{Schematic setup of a quantum dot tunnel coupled to one normal and two superconducting leads. The dot can be driven out of equilibrium by a bias voltage applied to the normal lead.}
\label{setup}
\end{center}
\end{figure}

We now turn our attention to the setup shown in Fig.~\ref{setup}.
As compared to the geometry considered so far, there is a third, normal (N),
lead with tunnel-coupling strength $\Gamma_{\text{N}}$, in addition to the two 
superconducting ones (L, R).
Again, we assume the same tunnel coupling $\Gamma_{\text{S}}$ and chemical
potential $\mu_{\text{S}}=0$ for both superconducting leads, and
$\Delta_{\text{L}}= \Delta_{\text{R}}^*=|\Delta|\exp(i\Phi/2)$.
The third lead allows for driving the quantum dot out of equilibrium by
applying a voltage between normal and superconducting leads, expressed
by a non vanishing chemical potential $\mu_{\text{N}}$ of the normal lead.
The quantities of interest are the Josephson current   
$J_{\text{jos}}=(J_{\text{L}}-J_{\text{R}})/2$ and the Andreev current in 
the normal lead $J_{\text{and}}=-(J_{\text{R}}+J_{\text{L}})$. 

In Ref.~\onlinecite{pala07} we studied this setup in the limit of weak
tunnel couplings.
We found that by applying a bias voltage between normal and superconducting
leads one can induce an out-of-equilibrium proximity effect in the quantum dot,
which, in turn, supports a Josephson coupling carried by first-order tunnel
processes instead of second order (cotunneling).
We described the non-equilibrium Josephson current as well as transitions from
$0$ to $\pi$-states perturbatively to first order in $\Gamma_{\text{S}}$.
This limited the applicability to a small range of gate voltages and
temperatures larger than the tunnel-coupling strengths.
The proximity effect was of purely non-equilibrium origin since
the influence of the superconducting leads on the quantum-dot spectrum,
typically associated with the picture of Andreev bound states, 
could not be resolved.

In the present paper, we want to go beyond the limit considered in Ref.~\onlinecite{pala07} 
for two reasons.
First, we aim at covering the full range of gate and bias voltages, thus,
including both equilibrium and non-equilibrium proximity effect.
Second, we are interested in mapping out the spectrum of Andreev bound
states of an interacting quantum dot.
To pursue both of these aims, we need to go beyond first-order transport in $\Gamma_{\text{S}}$.
As usual for interacting systems, the full problem for arbitrary
values of $|\Delta|$ and $\Gamma_{\text{S}}$ can only be solved 
approximatively.
In the limit of a large superconducting gap $|\Delta|\rightarrow\infty$, 
however, we are able to derive an exact result by resummation of the contributions
of all orders in $\Gamma_{\text{S}}$.
This is possible because for $|\Delta|\rightarrow\infty$ only a small subset 
of all diagrams contributes to the generalized rates:
the only superconducting lines that remain are anomalous ones that connect
vertices within one (the upper or the lower) propagator, with no other
vertex appearing in between.
This simplification is related to the inaccessibility of quasiparticle 
excitations in the superconducting leads and the fact that a Cooper pair 
should tunnel in a time interval $\propto \hbar /|\Delta|$, which 
becomes infinitesimal for $|\Delta|\rightarrow\infty$.
A rigorous proof is given in Appendix~\ref{rules_deriv}.

In the limit $|\Delta|\rightarrow\infty$ we can evaluate the current 
in the superconducting leads $\text{L}, \text{R}$ by means of Eq.~(\ref{jeta}),
i.e., we only need the pair amplitude of the quantum dot.
It is useful to introduce a dot isospin defined as \cite{pala07}   
\begin{equation}
I_x=\frac{P_0^{\text{D}}+P_{\text{D}}^0}{2}; \; I_y=i\frac{P_0^{\text{D}}-P_{\text{D}}^0}{2}; \; 
I_z=\frac{P_{\text{D}}-P_0}{2}.
\end{equation}
Finite $x$- and $y$-components of the isospin indicate coherent superpositions
of the dot being empty or doubly-occupied.

We rewrite the master equation for the dot reduced density matrix in the form of a Bloch equation for the isospin, taking into account all rates up to 
first order in $\Gamma_{\text{N}}$ [the order in $\Gamma_{\text{N}}$ is 
indicated by the superscript $(i)$ with $i=0,1$]. 
The Bloch equation for the isospin reads: 
\begin{equation}
  \label{bloch}
  0=\frac{d \mathbf I}{dt}= \mathbf{A} - \mathbf{R} \cdot \mathbf{I}
  + \mathbf{I}\times \mathbf{B} \, ,
\end{equation}
where the first, second, and third term describe generation, relaxation, and
rotation of the isospin, respectively.
The explicit expressions of the needed generalized rates are reported 
in Appendix~\ref{ratesDinf}.
In order to decouple the equations for the isospin from those for the diagonal probabilities we 
made use of the relations: $\displaystyle W_{\text{D}\sigma}^{(1)}+W_{\sigma\text{D}}^{(1)}-W_{\text{0}\sigma}^{(1)}-W_{\sigma\text{0}}^{(1)}=0$ and  $\displaystyle 2 W_{0\sigma}^{\text{D}\sigma(1)}-W_{00}^{\text{D}0(1)}
-W_{0\text{D}}^{\text{DD}(1)}=0$.  
The relaxation tensor
and the generation vector start in first order in $\Gamma_{\text{N}}$. 
The generation vector reads
\begin{equation}
  \mathbf{A}^{(1)}=\left( 
  \begin{array}{c} 
    \text{Re}\left\{W_{0\sigma}^{\text{D}\sigma(1)}\right\}\\
    -\text{Im}\left\{W_{0\sigma}^{\text{D}\sigma(1)}\right\}
    \\
    \frac{1}{2}\left(W_{\text{D}\sigma}^{(1)}-W_{0\sigma}^{(1)}\right)\\
  \end{array}
  \right).
\end{equation}
The only non vanishing elements of the relaxation tensor are:  $R_{xx}^{(1)}=R_{yy}^{(1)}=-\text{Re}\left\{W_{00}^{\text{DD}(1)}\right\}$, $R_{zz}^{(1)}=W_{\sigma0}^{(1)}+W_{\sigma\text{D}}^{(1)}$, and $R_{xz}^{(1)}=R_{zx}^{(1)}=\text{Re}\left\{W_{00}^{0\text{D}(1)}-W_{\text{D}0}^{\text{DD}(1)} \right\}$.
The effective magnetic field acting on the isospin has a zeroth-order component $\textbf{B}^{(0)}$ and a first-order component 
$\textbf{B}^{(1)}$, 
\begin{eqnarray}
\textbf{B}^{(0)}&=&\left(
\begin{array}{c}
 2 \Gamma_{\text{S}} \cos \Phi/2\\
0\\
-(2\epsilon+U)
\end{array}\right) \\
\textbf{B}^{(1)}&=&\left(
\begin{array}{c}
-\text{Im}\left\{W_{\text{D}0}^{\text{DD}(1)}-W_{00}^{0\text{D}(1)}\right\} \\
0\\
\text{Im}\left\{W_{00}^{\text{DD}(1)}\right\}
\end{array} 
\right).
\end{eqnarray}

The explicit expressions for the generation vector and the relaxation tensor can be written 
in a compact way, if we define the Andreev bound-state energies. These are given by the poles of the retarded Green's function of the dot for vanishing coupling to the normal lead,
\begin{equation}
  \label{abs}
  E_{\text{A},\gamma', \gamma} = 
  \gamma' \frac{U}{2}+\gamma \sqrt{\left(\epsilon+\frac{U}{2}\right)^2+
    \Gamma_{\text{S}}^2 \cos^2 \frac{\Phi}{2}},
\end{equation}
where $\gamma$ and $\gamma'$ can take the values $\pm 1$.
There are four resonances which lie pairwise around zero energy. 
We get for the generation vector: 
\begin{subequations}
\begin{eqnarray}
A_x^{(1)} &=&-\frac{\Gamma_{\text{S}}\Gamma_{\text{N}}}{4\ea}
\cos \frac{\Phi}{2} 
\sum_{\gamma,\gamma'=\pm}\gamma f_{\text{N}}(E_{\text{A},\gamma', \gamma})\\
A_y^{(1)} &=& 0\\
A_z^{(1)} &=& \frac{\Gamma_{\text{N}}}{4} \sum_{\gamma,\gamma'=\pm}\left(1+\gamma \frac{\epsilon+\frac{U}{2}}{\ea}\right)
\left[ f_{\text{N}}(E_{\text{A},\gamma', \gamma})-\frac{1}{2}\right] \, , 
\end{eqnarray}
\end{subequations}
with $\ea=\sqrt{\left(\epsilon+\frac{U}{2}\right)^2+\Gamma_{\text{S}}^2\cos^2 
\frac{\Phi}{2}}$, where the square-root dependence clearly indicates that 
the result is non-perturbative in $\Gamma_{\text{S}}$. 
The non-vanishing elements of the relaxation tensor are:
\begin{widetext}
\begin{subequations}
\begin{eqnarray}
  R_{xx}^{(1)}&=&R_{yy}^{(1)}= 
\frac{\Gamma_\text{N}}{2}\sum_{\gamma,\gamma'=\pm}
  \left(1-\gamma\frac{\epsilon+ U/2}{\ea}\right)\left[ \frac{1}{2}-\gamma' f_{\text{N}}(E_{\text{A},\gamma', \gamma}) \right]\\ 
  R_{zz}^{(1)}&=& \frac{\Gamma_{\text{N}}}{2} \sum_{\gamma,\gamma'=\pm}\left(1+\gamma \frac{\epsilon+U/2}{\ea}\right)
  \left[\frac{1}{2}-\gamma' f_{\text{N}}(E_{\text{A},\gamma',\gamma})\right]\\ 
  R_{xz}^{(1)}&=&R_{zx}^{(1)}=
  \frac{\Gamma_{\text{S}}\Gamma_{\text{N}}}{2\ea}\cos \frac{\Phi}{2} 
  \sum_{\gamma,\gamma'=\pm}\gamma\gamma' f_{\text{N}}(E_{\text{A},\gamma', \gamma})
\,.
\end{eqnarray}
\end{subequations}
\end{widetext}

By means of Eq.~(\ref{jeta}), the current in the superconducting leads can be written as 
$J_{\text{R,L}}=\frac{2e}{\hbar}\Gamma_{\text{S}}\left(I_y \cos \frac{\Phi}{2}\mp I_x \sin\frac{\Phi}{2}\right)$,
where the upper (lower) sign refers to the right (left) lead. 
Hence, the $x$- and $y$-component of the isospin provide the 
Josephson and Andreev currents, respectively,
\begin{eqnarray}
  J_{\text{jos}}&=&\frac{2e}{\hbar}\Gamma_{\text{S}} I_x \sin \frac{\Phi}{2}\\
  J_{\text{and}}&=&-\frac{4e}{\hbar}\Gamma_{\text{S}} I_y \cos \frac{\Phi}{2}\, ,
\end{eqnarray}
whereas the $z$-component is related to the charge in the quantum dot,
\begin{equation}
  Q=-e(1+2 I_z).
\end{equation}

We solve for the stationary solution for the isospin. 
Expanding the Eq.~(\ref{bloch}) to zeroth order in $\Gamma_{\text{N}}$ yields
$0=\mathbf{I}^{(0)} \times \mathbf{B}^{(0)}$, and, thus, 
$\mathbf{I}^{(0)} \parallel \mathbf{B}^{(0)}$.
To determine the proportionality constant, we multiply $\mathbf{B}^{(0)}$ from 
the left to Eq.~(\ref{bloch}) expanded to first order in $\Gamma_{\text{N}}$,
and obtain the zeroth-order result
\begin{equation}
  \mathbf{I}^{(0)}=\left(  \frac{\mathbf{A}^{(1)}\cdot \mathbf{B}^{(0)}} {\mathbf{B}^{(0)}\cdot\mathbf{R}^{(1)}\cdot \mathbf{B}^{(0)}}\right)\mathbf{B}^{(0)}
\, ,
\end{equation} 
which yields the Josephson current and the quantum-dot charge.
The Andreev current, on the other hand, is proportional to the $y$-component of the isospin and starts in first order in 
$\Gamma_{\text{N}}$. The first-order contribution to the $y$-component of the isospin can be derived by multiplying either $\mathbf{\hat{x}}$ or
$\mathbf{\hat{z}}$ from the left to Eq.~(\ref{bloch}) expanded to first order 
\begin{eqnarray}
  \label{iy1}
  I_y^{(1)} &=& 
  \frac{1}{B_x^{(0)}} \, \mathbf{\hat{z}}\cdot \left( \mathbf{A}^{(1)}
  - \mathbf{R}^{(1)}\cdot \mathbf{I}^{(0)} \right)
\\  
  &=& - \frac{1}{B_z^{(0)}} \, \mathbf{\hat{x}}\cdot \left( \mathbf{A}^{(1)}
  - \mathbf{R}^{(1)}\cdot \mathbf{I}^{(0)} \right)
\, .
\end{eqnarray}

The formation of a finite pair amplitude of the dot is favored if the empty and
doubly-occupied dot states are degenerate, $2\epsilon + U=0$.
In this case, however, the dot is preferably singly occupied in equilibrium, 
i.e., the proximity effect is strongly suppressed by Coulomb charging.
For finite values of the superconducting gap $|\Delta|$, a small Josephson
current through the dot can be established by cotunneling processes.
In the limit of infinite $|\Delta|$, however, this is not possible, and the
proximity effect and, thus, the Josephson current is exponentially suppressed.
In fact, we find for this regime $\mathbf{A}^{(1)} = 0$, i.e., no
isospin is generated.

The are two routes towards the generation of a finite dot pair amplitude.
One is to change the gate voltage such that empty or double occupation of the
dot becomes available.
Then, the tunnel coupling to the superconductors give rise to an equilibrium
proximity effect that, however, starts in higher order in the tunnel coupling
strength.
To achieve a finite pair amplitude at lowest order already, one has to apply a 
finite bias voltage at the normal lead.
This induces a non-equilibrium proximity effect that supports a first-order
Josephson current through the dot.

\subsubsection{Equilibrium}

First, we consider the equilibrium situation ($\mu_{\text{N}}=0$). 
In this case, the relation $\mathbf{\hat{x}}\cdot\mathbf{R}^{(1)}\cdot\mathbf{B}^{(0)}/\mathbf{\hat{z}}\cdot\mathbf{R}^{(1)}\cdot\mathbf{B}^{(0)}={A_x^{(1)}} /{A_z^{(1)}}$ ensures that no current flows in the normal lead.
The exact result for the equilibrium Josephson current in zeroth order in 
$\Gamma_{\text{N}}$ reads
\begin{widetext}
\begin{equation}
\label{jos_eq}
J_{\text{jos}}=\frac{e}{\hbar} 
\Gamma_{\text{S}}^2 \sin \Phi
\frac{\underset{\gamma,\gamma'=\pm}{\sum}\left(1+\gamma \frac{\epsilon+U/2}{\ea}\right)
\left[f(E_{\text{A},\gamma', \gamma})-\frac{1}{2}\right]
}
{ 
(2\epsilon+U)\underset{\gamma,\gamma'=\pm}{\sum}\left(1+\gamma \frac{\epsilon+U/2}{\ea}\right)
\left[\gamma'f(E_{\text{A},\gamma', \gamma})-\frac{1}{2}\right]
+\frac{2}{\ea}\Gamma_{\text{S}}^2 \cos^2(\Phi/2)
\underset{\gamma,\gamma'=\pm}{\sum}\gamma\gamma'f(E_{\text{A},\gamma', \gamma})
}\, ,
\end{equation}
\end{widetext} 
where $f(\omega)$ is the Fermi function with zero chemical potential.
Notice that the only role played by the normal lead is to provide a mechanism for the electrons in the dot to reach equilibrium. Expanding Eq.~(\ref{jos_eq}) to second order in $\Gamma_{\text{S}}$ we recover the result of Eq.~(\ref{jos2Dinf}). 

\begin{figure}
\begin{center}
\includegraphics[width=3.1in]{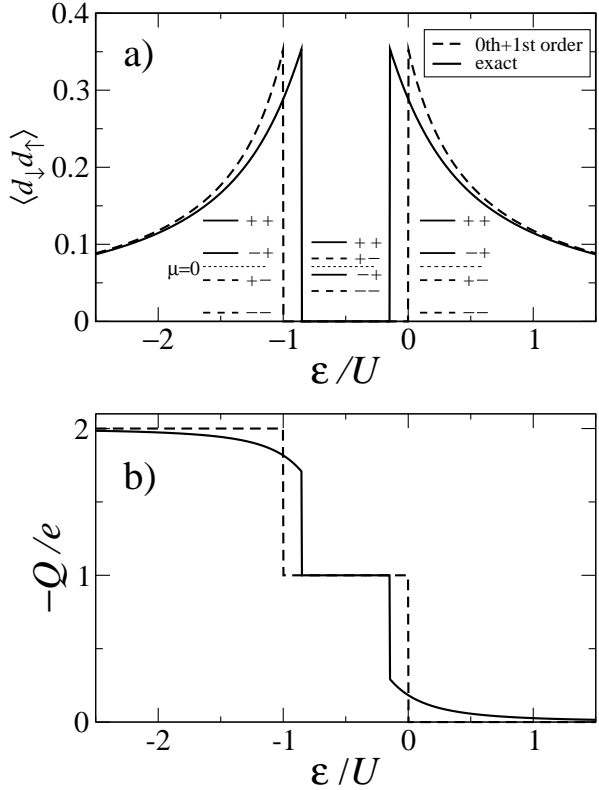}\\
\caption{
Dot pair amplitude $\langle d_{\downarrow}d_{\uparrow}\rangle$ (a) and charge (b) as a function of the level position $\epsilon$ at zero temperature. 
Note that the pair amplitude in equilibrium for the symmetric setup is real. The dashed lines refer to the pair amplitude and charge calculated up to first-order in $\Gamma_{\text{S}}$;  
for the pair amplitude the $0$th-order contribution vanishes, while for the charge the first-order 
one does.  
In the inset of panel a) we show a schematic picture of the Andreev bound state energies $E_{\text{A},\gamma',\gamma}$ in the three different regions, where the values $\gamma'\ \gamma$ are indicated next to the level. The sign of the contributions of the bound states $E_{\text{A},\gamma',\gamma}$ to $\langle d_{\downarrow}d_{\uparrow}\rangle$ is determined by the index $\gamma$. We have depicted the levels with $\gamma=+$ by a solid line and those with $\gamma=-$ by a dashed line. 
The other parameters used in the simulation are: $\Gamma_{\text{S}}/U=0.5$, and $\Phi=\pi/2$.}
\label{figeq}
\end{center} 
\end{figure}

In equilibrium, the pair amplitude of the quantum dot $\langle d_{\downarrow}d_{\uparrow}\rangle$ for the symmetric setup and symmetric gauge is real, i.e. $\langle d_{\downarrow}d_{\uparrow}\rangle=I_x$ since $I_y=0$. 
In panel (a) of Fig.~\ref{figeq} we plot the pair amplitude as a function of the level position $\epsilon$ for zero temperature. In particular we note that a quantum phase transition occurs if $U/2>\Gamma_{\text{S}} |\cos (\Phi/2)|$ 
at $\epsilon=\bar{\epsilon}_\pm=-U/2\pm\sqrt{(U/2)^2-\Gamma^2_{\text{S}} \cos^2 (\Phi/2)}$. The values $\bar{\epsilon}_\pm$ where the transition takes place depend on the tunnel coupling $\Gamma_{\text{S}}$ and on the superconducting phase difference. The  pair amplitude in first order $\Gamma_{\text{S}}$, see Eq.~(\ref{p0D}), which gives rise to the second-order Josephson current,  exhibits the phase transition at different values of the level position. The behavior of the pair amplitude can be understood by considering the Andreev-bound-state configuration in each region (see inset of panel (a) of Fig.~\ref{figeq}). 
Panel (b) of  Fig.~\ref{figeq} shows the charge of the dot as a function of the level position. In this case, the first-order $\Gamma_{\text{S}}$ correction vanishes. The full result for the charge shows that due to proximity effect the charge on the dot is not always quantized.~\cite{blatter07}

\subsubsection{Non-equilibrium}
\begin{figure}
\begin{center}
\includegraphics[width=3.1in]{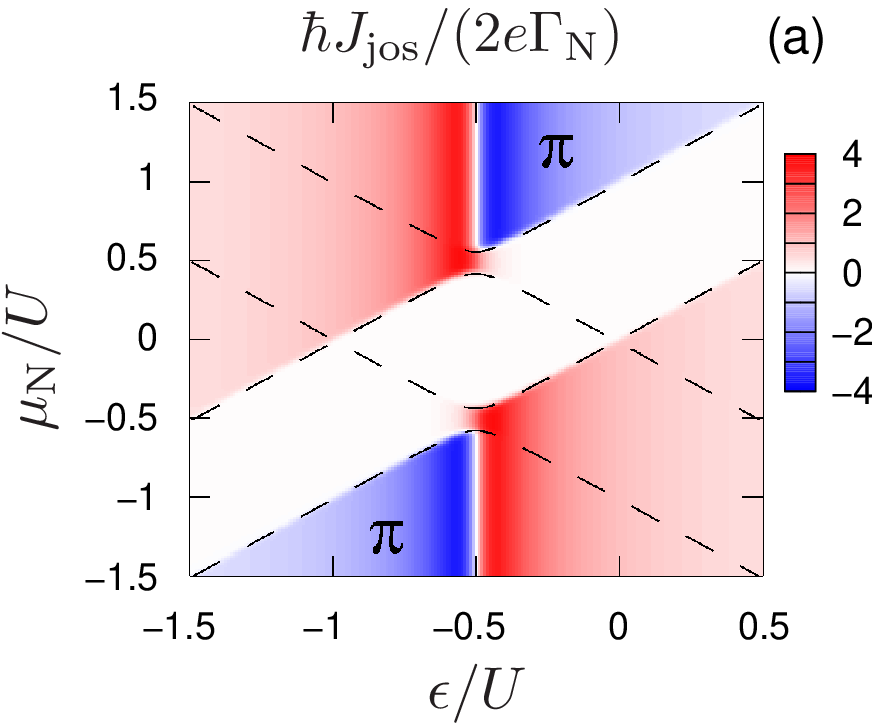}\\
\includegraphics[width=3.1in]{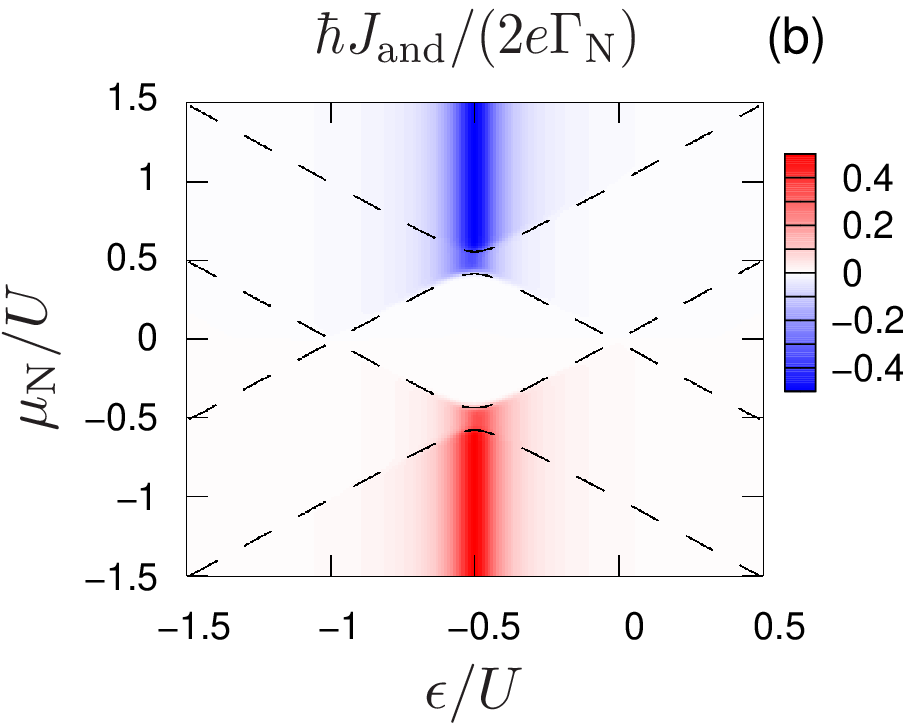}\\
\includegraphics[width=3.1in]{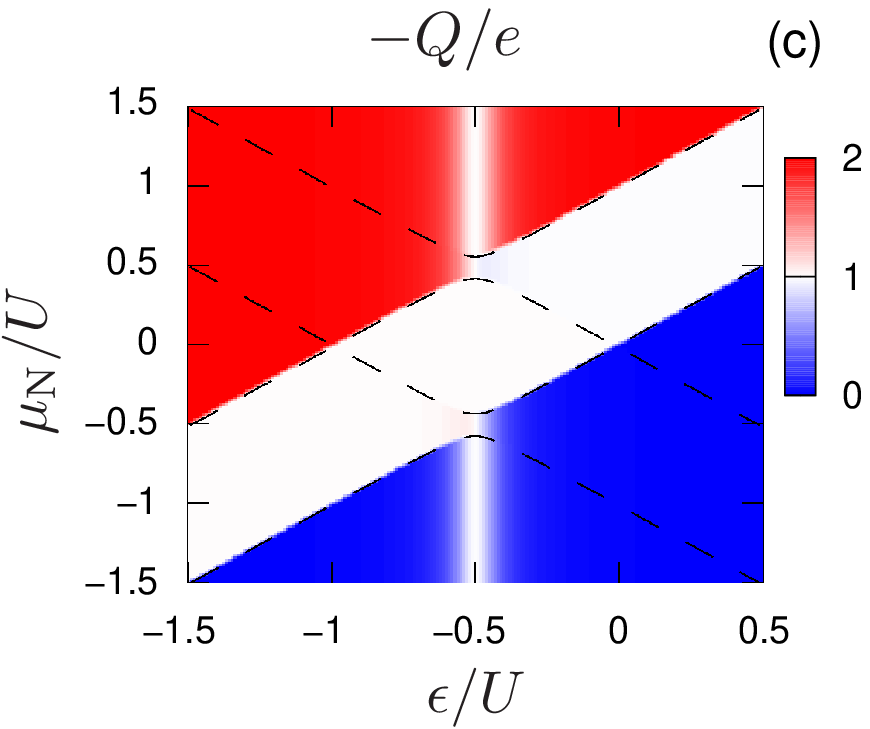}\\
\caption{(Color online) Density plot of the Josephson current (a), the Andreev current (b), and the charge on the quantum dot (c) as a function of the level position $\epsilon$ and of the chemical potential $\mu_{\text{N}}$. In panel (a), the region where the system behaves as 
a $\pi$-junction is indicated by the symbol $\pi$. The dashed lines map the Andreev bound states: 
$\mu_{\text{N}}=E_{\text{A},\gamma',\gamma}$.
The other parameters used in the simulation are: $k_{\text{B}} T/U=0.01$, $\Gamma_{\text{N}}/U=0.005$, 
$\Gamma_{\text{S}}/U=0.1$, and $\Phi=\pi/2$.}
\label{figinf0.1}
\end{center} 
\end{figure}

\begin{figure}
\begin{center}
\includegraphics[width=3.1in]{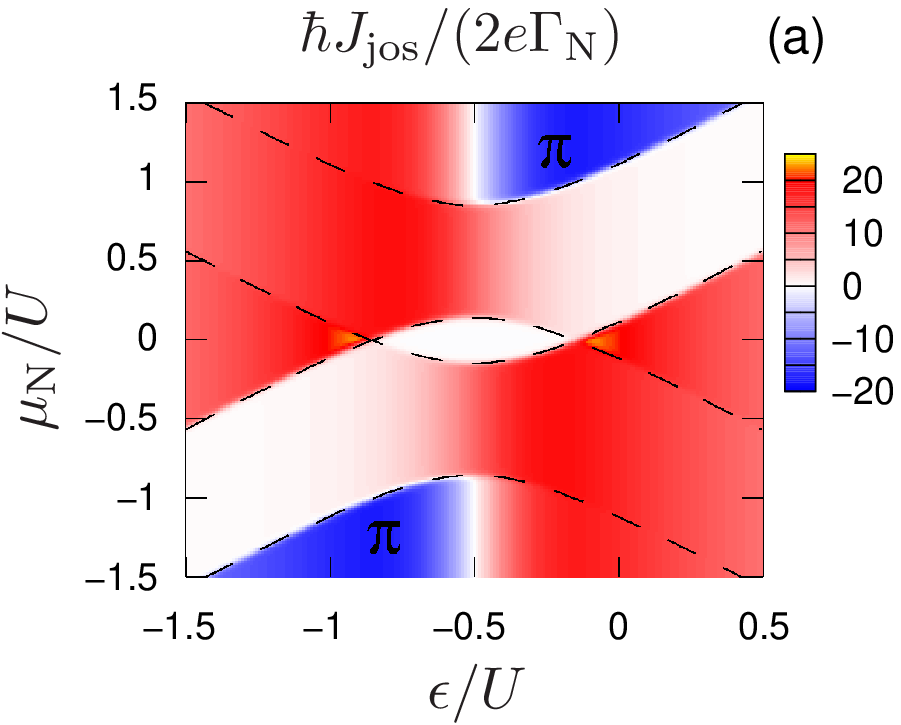}\\
\includegraphics[width=3.1in]{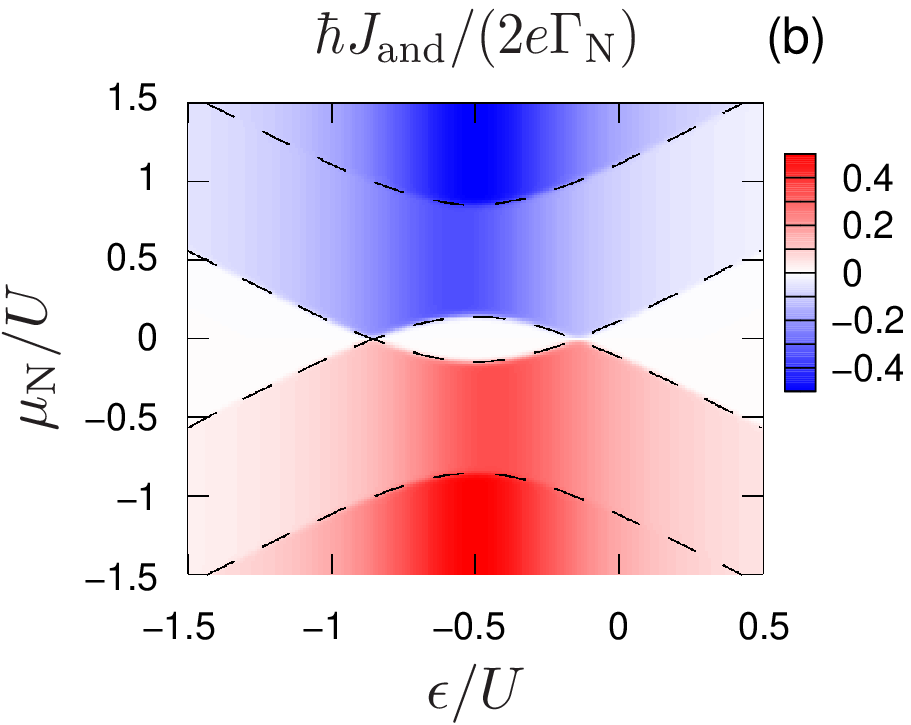}\\
\includegraphics[width=3.1in]{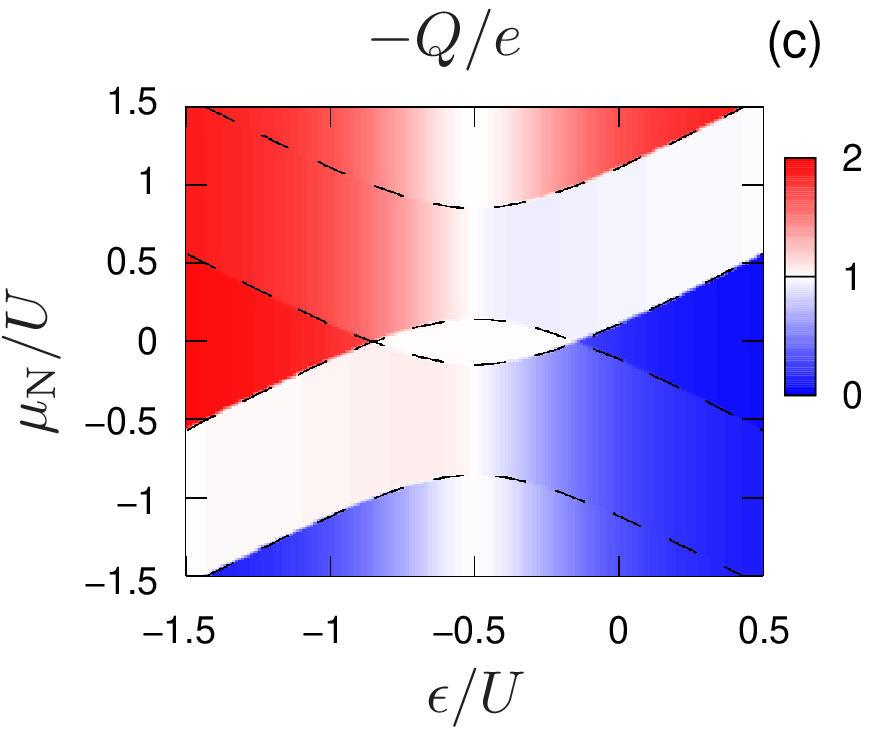}\\
\caption{(Color online) Density plot of the Josephson current (a), the Andreev current (b), and the charge on the quantum dot (c) as a function of the level position $\epsilon$ and of the chemical potential $\mu_{\text{N}}$. In panel (a), the region where the system behaves as 
a $\pi$-junction is indicated by the symbol $\pi$. The dashed lines map the Andreev bound states: 
$\mu_{\text{N}}=E_{\text{A},\gamma',\gamma}$.
The other parameters used in the simulation are: $k_{\text{B}} T/U=0.01$, $\Gamma_{\text{N}}/U=0.005$, 
$\Gamma_{\text{S}}/U=0.5$, and $\Phi=\pi/2$.}
\label{figinf0.5}
\end{center} 
\end{figure}

\begin{figure}
\begin{center}
\includegraphics[width=3.1in]{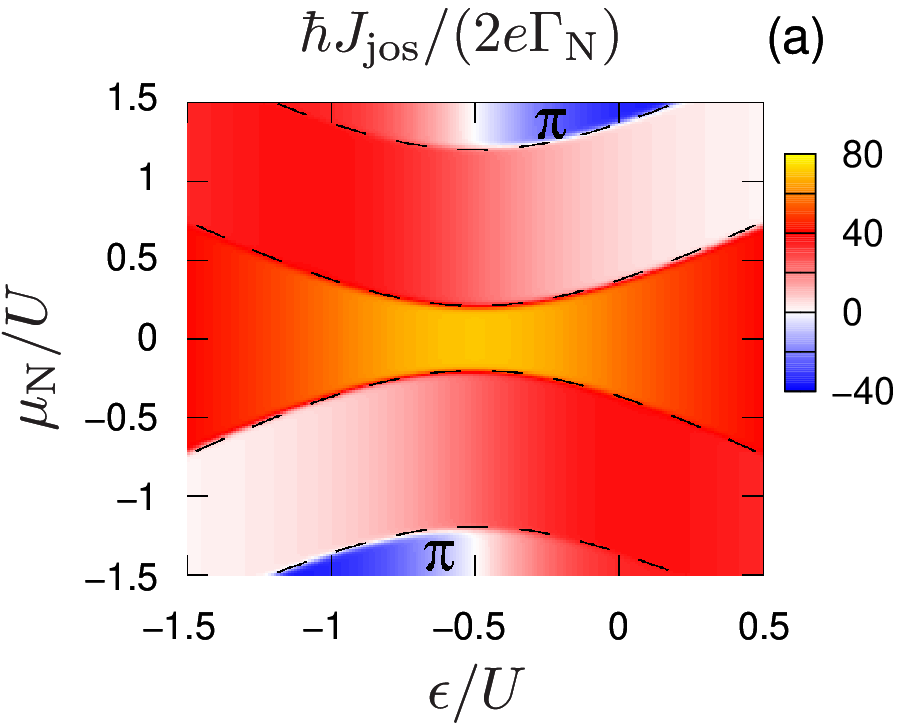}\\
\includegraphics[width=3.1in]{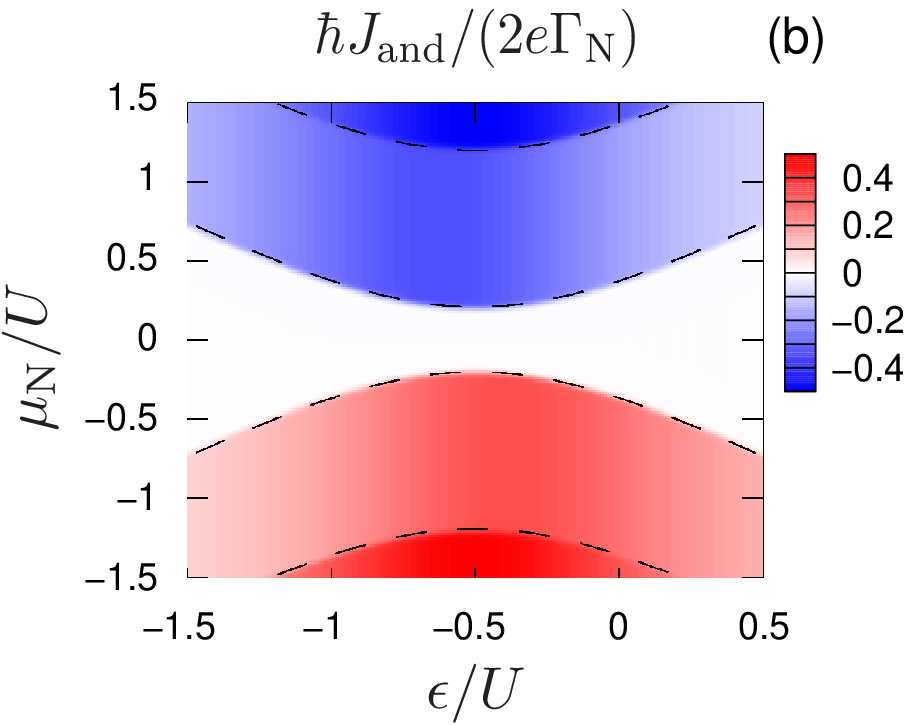}\\
\includegraphics[width=3.1in]{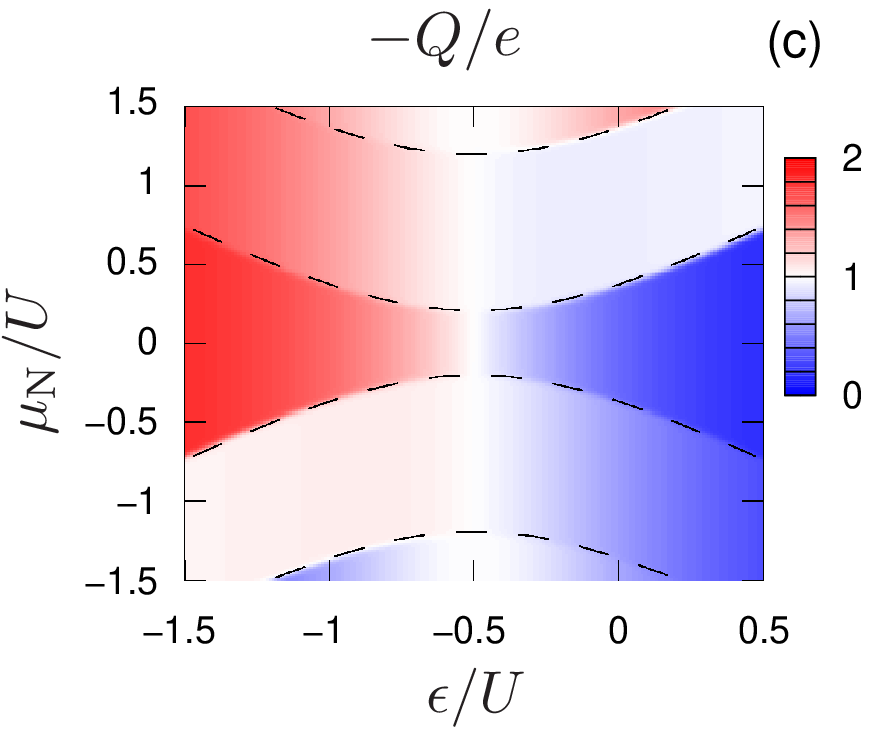}\\
\caption{(Color online) Density plot of the Josephson current (a), the Andreev current (b), and the charge on the quantum dot (c) as a function of the level position $\epsilon$ and of the chemical potential $\mu_{\text{N}}$. In panel (a), the region where the system behaves as 
a $\pi$-junction is indicated by the symbol $\pi$. The dashed lines map the Andreev bound states: 
$\mu_{\text{N}}=E_{\text{A},\gamma',\gamma}$.
The other parameters used in the simulation are: $k_{\text{B}} T/U=0.01$, $\Gamma_{\text{N}}/U=0.005$, 
$\Gamma_{\text{S}}/U=1$, and $\Phi=\pi/2$.}
\label{figinf1.0}
\end{center} 
\end{figure}

Next we turn our attention to the non-equilibrium situation ($\mu_{\text{N}}\ne 0$). Applying a bias voltage 
to the normal lead produces a finite current in $\text{N}$, which is sustained by Andreev-reflection 
processes. 
We do not give here the explicit analytical expressions for the zeroth-order $J_{\text{jos}}$ and 
the first-order $J_{\text{and}}$ since they are rather lengthy.
Instead, in Figs.~\ref{figinf0.1}, \ref{figinf0.5}, \ref{figinf1.0} we plot the Josephson current, the 
Andreev current, and the dot charge as a function of the level position $\epsilon$ and of the chemical potential of the normal lead $\mu_{\text{N}}$, for different values of the tunnel-coupling with the superconductor.

In Fig.~\ref{figinf0.1}(a) one can see how the Josephson current can be controlled by the chemical potential of 
the normal lead. 
There is, first, a broad region set by the charging energy in which the 
Josephson current is suppressed.
Second, a $\pi$-transition can be driven both by the transport voltage and 
by the gate voltage controlling the level position. 
For fixed $\mu_{\text{N}}$ the transition occurs at $\epsilon=-U/2$, 
i.e. when the energy of the empty and double occupied dot are degenerate. 
We remark that this transition is slightly shifted when higher-order 
corrections to the effective field are included. 
In fact, near the transition $B_z^{(0)}=2\epsilon+U$ becomes small and hence $B_z^{(1)}$ needs to be 
taken into account. This has been done in Ref.~\onlinecite{pala07} in the weak-proximity limit. 
Panel (b) of Fig.~\ref{figinf0.1} shows the Andreev current.
It is largest at $\epsilon=-U/2$ outside the region where charging energy
suppresses transport.
Panel (c) of Fig.~\ref{figinf0.1} shows the dot charge.
We find a pronounced feature around $\epsilon=-U/2$ that is associated 
with generating a $y$-component of the isospin by rotation out of the 
$z$-direction.
 
In Figs.~\ref{figinf0.5} and  \ref{figinf1.0}, the coupling to the superconducting lead is stronger, comparable to the Coulomb interaction strength, and the term proportional to $\Gamma_{\text{S}}$ in  $\sqrt{(\epsilon+U/2)^2+\Gamma_{\text{S}}^2 \cos^2 (\Phi/2)}$ becomes more important, leading to 
a more pronounced splitting of the Andreev bound-state energies. 
We stress here that the current in the normal lead as a function of both gate and transport voltage maps the energies of the Andreev bound states in the dot. Therefore, measuring the current in the normal lead allows to perform a \textit{an Andreev-bound-state spectroscopy} and, hence, to gather information on the superconducting correlations induced in the dot.

\section{Conclusion}
\label{s7}
We have presented a real-time diagrammatic transport theory for systems composed of interacting  quantum dots coupled both to normal and superconducting leads. First, we have applied this theory to 
study the Josephson current through a quantum-dot tunnel coupled to two superconductors in second order in the tunnel-coupling strengths. In particular,  we have studied how a $\pi$-phase develops with increasing on-site Coulomb repulsion.  Next, we have considered a quantum dot coupled to one normal and two superconducting leads, in the limit of large superconducting gap. In this regime, all orders 
in the tunnel-coupling strengths with the superconductors can be summed. This enabled us to investigate the strong-proximity regime. In particular, 
we analyze the Josephson current and identify the parameter regions where the system behaves as $\pi$-junction; the $\pi$-transition can be triggered both by the dot level position and the bias voltage.   We find also that a spectroscopy of the Andreev bound states of the system can be realized by measuring the Josephson current between tho two superconductors, the Andreev current in the normal lead or the charge of the dot as a function of both the dot level-position and the bias voltage.

\begin{acknowledgments}
We would like to thank W. Belzig, F. S. Bergeret, R. Fazio, A. Shnirman, and A. Volkov for useful discussions.
Financial support from the DFG via SFB 491 is acknowledged.
\end{acknowledgments}

\begin{appendix}

\section{Second-order, finite $|\Delta|$}
\label{appsecond}
\begin{figure}
\begin{center}
\includegraphics[width=3.1in]{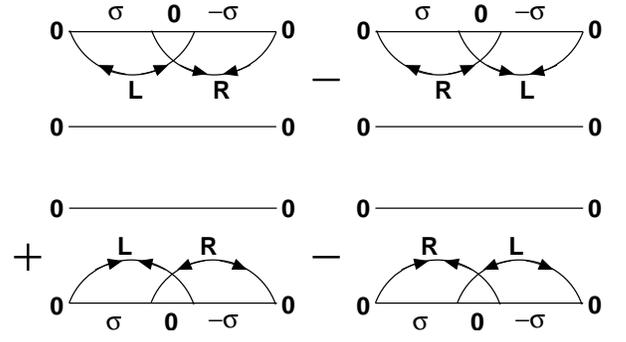}\\
\caption{Second-order diagrams contributing to the current rate $W_{00}^{00\text{L}(2)}$.}
\label{currentrate}
\end{center} 
\end{figure}

In this Appendix we show, as an example, the calculation of  the second-order current rate $W_{00}^{00\text{L}(2)}$. 
The second-order diagrams contributing to this rate are shown in 
Fig.~\ref{currentrate}; the signs have been assigned making use of Rule 6 in Section \ref{ss4}.
Using the diagrammatic rules of Section \ref{ss4} we get 
\begin{widetext}
\begin{eqnarray*}
W_{00}^{00\text{L}(2)}&=&-8\Gamma_{\text{S}}^2 \sin \Phi\int_{|\Delta|}^{\infty} \frac{d \omega}{2\pi}
\frac{|\Delta|}{\sqrt{\omega^2-|\Delta|^2}}\int_{|\Delta|}^{\infty} \frac{d \omega'}{2\pi}
\frac{|\Delta|}{\sqrt{{\omega'}^2-|\Delta|^2}}\frac{1}{\omega+\epsilon}\frac{1}{\omega+\omega'}
\frac{1}{\omega'+\epsilon}\\
&=&-2 \frac{\Gamma_{\text{S}}^2}{|\Delta| } \sin \Phi F\left(\frac{\epsilon}{|\Delta|}\right)
\end{eqnarray*}
\end{widetext}

\section{Derivation of the rules for $|\Delta|\rightarrow\infty$}
\label{rules_deriv}
Here, we give a rigorous proof of the rules which in the $|\Delta|\rightarrow\infty$ limit 
allow us to greatly reduce the number of diagrams to be considered.

Rule (i): \textit{No vertex should be considered between the two vertices of a line with a superconducting lead.} \\
Le us consider a diagram where a vertex $v'$ exists between the two vertices of a 
superconducting  line with energy $\omega$. Let the vertex $v'$ be associated with a line with energy $\omega'$.  According to the diagrammatic rules 2 and 3, this diagram contains the factor\cite{normalline} 
$\frac{1}{\pm\omega \dots} \frac{1}{\pm\omega \pm\omega'\dots} D(\omega)|\Delta/\omega|$ which upon 
integration over $\omega$ vanishes as $1/|\Delta|$. On the other hand, if no vertex is inserted between the two  vertices of the superconducting line, the diagram contains the factor\cite{normalline} 
$\frac{1}{\pm\omega \dots} D(\omega)|\Delta/\omega|$ 
which remains finite upon integration over $\omega$. 

Rule (ii): \textit{No line with a superconductor joining the upper and lower propagator should be  considered.}\\ 
Let us consider the diagram where a superconducting line is running from the upper to the lower  propagator and the vertex on the upper propagator is on the left of 
the one on the lower propagator. In virtue of rule (i), the diagram with the two vertices  swapped, i.e. with the vertex on the upper propagator being on the right of the one on the lower  propagator, also exists. These two diagrams cancels each other for $|\Delta|\rightarrow \infty$. 

\begin{figure}
\begin{center}
\includegraphics[width=3.1in]{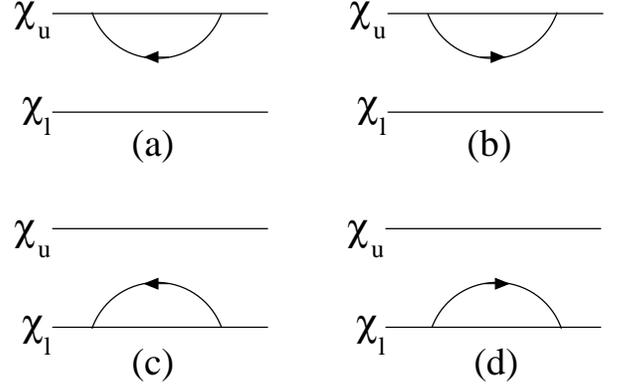}\\
\caption{Possible insertion of a normal line in the upper or in the lower propagator. Note that these insertions are a part of a larger irreducible diagram and hence there are other lines 
(not shown) running from one side to the other of the insertion.}
\label{normalinsert}
\end{center} 
\end{figure}

Rule (iii): \textit{No normal line with a superconductor should be considered.}\\
Let us consider a part of a diagram with a state $|\chi_{\text{u}}\rangle$ running on the upper part of the Keldysh contour and 
with a state $|\chi_{\text{l}}\rangle$ on the lower part. In virtue of the two 
previous rules, there are only four possible ways of inserting a normal line with a superconducting  lead, which are schematically depicted in Fig.~\ref{normalinsert}. 
In the large-gap limit, 
the diagrams corresponding to the possible insertion of a normal line, have the same absolute value.  But the diagrams arising from the insertion in the lower propagator (shown in the second line of 
Fig. \ref{normalinsert}) have an opposite sign with respect to the ones in the upper propagator (first 
line of Fig. \ref{normalinsert}). Finally, it easy to prove that for any $|\chi_{\text{u}}\rangle$ 
and $|\chi_{\text{l}}\rangle$ there are, for our single-level model, only four possible insertion: two in the upper propagator and two in the lower propagator. Hence, the sum of all these diagrams vanishes. To clarify this point, let us consider the exemplary case that 
$|\chi_{\text{u}}\rangle= |0\rangle$ and  $|\chi_{\text{l}}\rangle = |\uparrow\rangle$: then 
the possible insertions are: (a) with intermediate state $|\uparrow\rangle$ or $|\downarrow\rangle$ for the upper propagator; (c) with intermediate 
state $|0\rangle$ and (d) with intermediate state $|\text{D}\rangle$ for the lower propagator. 

\section{Calculation of a generalized rate in the $|\Delta|\rightarrow\infty$ limit}

\begin{figure}
\begin{center}
\includegraphics[width=2.7in]{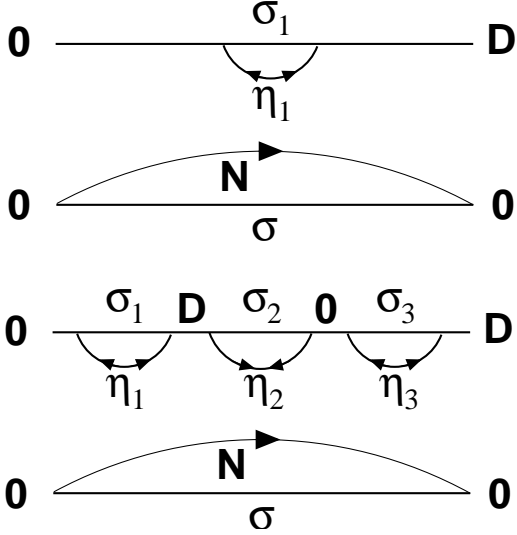}\\
\caption{First two contributions to $W_{00}^{\text{D}0(1)}$.}
\label{diaginf}
\end{center} 
\end{figure}

Here, we show in one example how all contributions in $\Gamma_{\text{S}}$ can be  summed up. 
We consider the off-diagonal rate $W_{00}^{\text{D}0(1)}$ in first order in $\Gamma_{\text{N}}$ and we add all contributions in $\Gamma_{\text{S}}$. In particular, only diagrams with an odd number of  anomalous lines on the upper propagator contribute to this rate. The first two diagrams are shown in 
Fig.~\ref{diaginf}. 
The contribution with $2n+1$ anomalous line reads
\begin{widetext}
\begin{equation*}
2i \Gamma_{\text{N}}\int \frac{d\omega}{2\pi}
f_{\text{N}}(\omega)\left(\frac{1}{-\omega+\epsilon+i
    0^{+}}\,\cdot \,\frac{1}{-\omega+\epsilon-U+i0^{+}}\right)^{n+1}
\left(\Gamma_{\text{S}}\cos \frac{\Phi}{2} \right)^{2n+1}.
\end{equation*} 
Summing up all terms we get 
\begin{eqnarray*}
W_{00}^{\text{D}0(1)}&=&2i \Gamma_{\text{N}}\Gamma_{\text{S}}\cos\frac{\Phi}{2} \int \frac{d\omega}{2\pi} f_{\text{N}}(\omega) 
\frac{1}{(\omega-\epsilon-i 0^{+})(\omega-\epsilon+U-i0^{+})-\left(\Gamma_{\text{S}}\cos(\Phi/2) \right)^{2}}\\
& =&i \frac{\Gamma_{\text{N}}\Gamma_{\text{S}}}{\ea} \cos\frac{\Phi}{2}
\int \frac{d\omega}{2\pi} f_{\text{N}}(\omega) \left( \frac{1}{\omega+U/2-\ea-i0^{+}}- \frac{1}{\omega+U/2+\ea-i0^{+}}
\right),
\end{eqnarray*}
\end{widetext} 
with $\ea=\sqrt{(\epsilon+U/2)^2+\Gamma^2_{\text{S}} \cos^2(\Phi/2)}$.

\section{Generalized rates to all orders in $\Gamma_\text{S}$}
\label{ratesDinf}
In this Appendix we give the expression for the generalized rates, which are necessary to compute the Josephson current in zeroth-order $\Gamma_{\text{N}}$ and the Andreev current in first order. 
The diagonal rates start in first-order $\Gamma_{\text{N}}$ and they are given by 
\begin{eqnarray*}
W_{0\sigma}^{(1)}&=&\frac{\Gamma_{\text{N}}}{2}\sum_{\gamma=\pm} \left(1+\gamma\frac{\epsilon+ U/2}{\ea}\right) 
\left[ 1-f_{\text{N}}(E_{\text{A},-, \gamma}) \right]\\
W_{\sigma0}^{(1)}&=&\frac{\Gamma_{\text{N}}}{2} \sum_{\gamma=\pm}\left(1+\gamma\frac{\epsilon+ U/2}{\ea}\right) 
f_{\text{N}}(E_{\text{A},-, \gamma})\\
W_{\sigma\text{D}}^{(1)}&=&\frac{\Gamma_{\text{N}}}{2} \sum_{\gamma=\pm} \left(1+\gamma\frac{\epsilon+ U/2}{\ea}\right) \left[ 1-f_{\text{N}}(E_{\text{A},+, \gamma}) \right]\\
W_{\text{D}\sigma}^{(1)}&=&\frac{\Gamma_{\text{N}}}{2}\sum_{\gamma=\pm} \left(1+\gamma\frac{\epsilon+ U/2}{\ea}\right) f_{\text{N}}(E_{\text{A},+, \gamma}), 
\end{eqnarray*}
where the Andreev bound-state energies read 
\begin{equation*}
  E_{\text{A},\gamma', \gamma} = 
  \gamma' \frac{U}{2}+\gamma \sqrt{\left(\epsilon+\frac{U}{2}\right)^2+
    \Gamma_{\text{S}}^2 \cos^2 \frac{\Phi}{2}}.
\end{equation*}

Some of the off-diagonal rates start in zeroth-order $\Gamma_{\text{N}}$. In particular, we have 
$W_{00}^{\text{D}0(0)}=(W_{\text{D}0}^{00(0)})^*=
W_{00}^{0\text{D}(0)}=(W_{0\text{D}}^{00(0)})^*=(W_{\text{0D}}^{\text{DD}(0)})^*=W_{\text{DD}}^{\text{0D}(0)}=(W_{\text{D}0}^{\text{DD}(0)})^*=W_{\text{DD}}^{\text{D0}(0)}=i \Gamma_{\text{S}} \cos\Phi/2$.
We also need the real part of the first-order corrections to these rates. Notice that the following relations hold
$\text{Re}\left\{ W_{00}^{\text{D}0(1)}\right\}=\text{Re}\left\{ W_{\text{D}0}^{00(1)}\right\}=
\text{Re}\left\{ W_{00}^{0\text{D}(1)}\right\}=\text{Re}\left\{ W_{0\text{D}}^{00(1)}\right\}$ and 
$\text{Re}\left\{ W_{\text{0D}}^{\text{DD}(1)}\right\}=\text{Re}\left\{ W_{\text{DD}}^{\text{0D}(1)}\right\}=\text{Re}\left\{ W_{\text{D}0}^{\text{DD}(1)}\right\}=\text{Re}\left\{ W_{\text{DD}}^{\text{D0}(1)}\right\}$. The first-order corrections read
\begin{eqnarray*}
\text{Re}\left\{ W_{00}^{\text{D}0(1)}\right\}
&=&-\Gamma_{\text{S}} \cos\frac{\Phi}{2}\frac{\Gamma_\text{N}}{2\ea} \sum_{\gamma=\pm} \gamma f_{\text{N}}(E_{\text{A},-, \gamma})\\
\text{Re}\left\{ W_{\text{0D}}^{\text{DD}(1)}\right\} &=& -\Gamma_{\text{S}} \cos\frac{\Phi}{2}\frac{\Gamma_\text{N}}{2\ea} \sum_{\gamma=\pm} \gamma f_{\text{N}}(E_{\text{A},+, \gamma}).
\end{eqnarray*}
There are also some off-diagonal rates which start in first order in $\Gamma_{\text{N}}$:
\begin{widetext}
\begin{eqnarray*}
\text{Re}\left\{ W_{0 \sigma}^{\text{D}\sigma(1)}\right\}&=&-\text{Re}\left\{ W_{\sigma 0}^{\sigma \text{D}(1)}\right\}=
-\frac{\Gamma_{\text{S}}\Gamma_\text{N}}{4\ea}  \cos\frac{\Phi}{2} \sum_{\gamma,\gamma'=\pm} \gamma f_{\text{N}}(E_{\text{A},\gamma', \gamma}) \\
\text{Re}\left\{ W_{0 0}^{\text{DD}(1)}\right\}&=&\text{Re}\left\{ W_{\text{DD}}^{\text{00}(1)}\right\}=
\frac{\Gamma_\text{N}}{2}\sum_{\gamma,\gamma'=\pm}
\left(1-\gamma\frac{\epsilon+ U/2}{\ea}\right)\left[\gamma' f_{\text{N}}(E_{\text{A},\gamma', \gamma}) -\frac{1}{2} \right].
\end{eqnarray*}
\end{widetext}
\end{appendix}



\begin{thebibliography}{99}

\bibitem{nanotubes}
M. R. Buitelaar, T. Nussbaumer, and C. Sch\"onenberger, Phys. Rev. Lett. \textbf{89}, 256801 (2002);
J.-P. Cleuziou, W. Wernsdorfer, V. Bouchiat, T. Ondar\c{c}uhu, and M. Monthioux,
Nature Nanotechnology \textbf{1}, 53 (2006);
P. Jarillo-Herrero, J. A. van Dam, and L. P. Kouwenhoven,
Nature \textbf{439}, 953 (2006);
H. I. J{\o}rgensen, K. Grove-Rasmussen, T. Novotn\'y, K. Flensberg, and P. E. Lindelof, Phys. Rev. Lett. \textbf{96}, 207003 (2006).

\bibitem{nanowires}
J.A.~van Dam, Y.V.~Nazarov, E.P.A.M.~Bakkers, S.~De Franceschi, 
and L.P.~Kouwenhoven, Nature {\bf 442}, 667 (2006); 
T. Sand-Jespersen, J. Paaske, B. M. Andersen, K. Grove-Rasmussen, H. I. J{\o}rgensen, M. Aagesen, C. S{\o}rensen, P. E. Lindelof, K. Flensberg, and J. Nyg{\aa}rd, Phys. Rev. Lett. \textbf{99}, 126603 (2007).

\bibitem{buizert07} C.~Buizert, A.~Oiwa, K.~Shibata, K.~Hirakawa, and S. Tarucha, 
Phys. Rev. Lett. \textbf{99}, 136806 (2007). 

\bibitem{fazio98} R. Fazio and R. Raimondi, Phys. Rev. Lett. {\bf 80}, 2913
        (1998); Phys. Rev. Lett. {\bf 82}, 4950 (1999).

\bibitem{kang98} K. Kang, Phys. Rev. B {\bf 58}, 9641 (1998).

\bibitem{schwab99} P. Schwab and R. Raimondi, Phys. Rev. B {\bf 59}, 1637
        (1999).

\bibitem{clerk00} A. A. Clerk, V. Ambegaokar, and S. Hershfield, Phys. Rev. B
        {\bf 61}, 3555 (2000).


\bibitem{lambert00} S. Shapira, E. H. Linfield, C. J. Lambert, R. Seviour,
        A. F. Volkov, and A. V. Zaitsev, Phys. Rev. Lett. {\bf 84}, 159 (2000).

\bibitem{cuevas01} J. C. Cuevas, A. Levy Yeyati, and A. Mart\'in-Rodero,
        Phys. Rev. B {\bf 63}, 094515 (2001).

\bibitem{beenakker92} C. W. J. Beenakker and H. van Houten, 
in \textit{Single-Electron Tunneling and Mesoscopic Devices}, edited by H. Koch and H. L\"ubbig, Springer, Berlin, 1992, pp. 175-179. 

\bibitem{glazman89} L. I. Glazman and K. A. Matveev, JETP Lett. \textbf{49}, 659 (1989).

\bibitem{spivak91} B. I. Spivak and S. A. Kivelson, Phys. Rev. B \textbf{43}, 3740 (1991).

\bibitem{rozhkov01}
A.V.~Rozhkov, D.P.~Arovas, and F.~Guinea, 
Phys.~Rev.~B {\bf 64}, 233301 (2001).

\bibitem{clerk00bis}
A. A.~Clerk and V. Ambegaokar, Phys. Rev. B {\bf 61}, 9109 (2000).

\bibitem{avishai03}
Y.~Avishai, A.~Golub, and A.D.~Zaikin, 
Phys. Rev. B {\bf 67}, 041301(R) (2003).

\bibitem{sellier05} G. Sellier, T. Kopp, J. Kroha, and Y. S. Barash, Phys. Rev. B {\bf 72}, 174502 (2005).

\bibitem{lopez07}  R. L\'opez, Mahn-Soo Choi, and R. Aguado, Phys. Rev. B {\bf 75}, 045132 (2007).

\bibitem{bergeret06}
F. S. Bergeret, A. Levy Yeyati, and A. Martin-Rodero, Phys. Rev. B \textbf{74}, 132505 (2006).

\bibitem{karrasch07}
C. Karrasch, A. Oguri, and V. Meden, Phys. Rev. B \textbf{77}, 024517 (2008).

\bibitem{nussinov05} Z. Nussinov, A. Shnirman, D. P. Arovas, A. V. Balatsky, and J. X. Zhu,
 Phys. Rev. B \textbf{71}, 214520 (2005).

\bibitem{cuevas97} 
A. Levy Yeyati, J. C. Cuevas, A. L\'opez-D\'avalos, and A. Mart\'in-Rodero, Phys. Rev. B \textbf{55}, R6137 (1997).

\bibitem{johansson99} 
G. Johansson, E. N. Bratus, V. S. Shumeiko, and G. Wendin, 
Phys. Rev. B \textbf{60}, 1382 (1999).

\bibitem{blatter07}
I. A. Sadovskyy, G. B. Lesovik, and G. Blatter, Phys. Rev. B \textbf{75}, 195334 (2007).

\bibitem{ando95} 
S. Ishizaka, J Sone, and T. Ando, Phys. Rev. B \textbf{52}, 8358 (1995). 

\bibitem{choi04} 
Mahn-Soo Choi, Minchul Lee,  K. Kang, and W. Belzig, Phys. Rev. B \textbf{70}, 020502(R) 
(2004).

\bibitem{siano04} 
F. Siano and R. Egger, Phys. Rev. Lett. \textbf{93}, 047002 (2004). 

\bibitem{vecino03} 
E. Vecino, A. Mart\'in-Rodero, and A. Levy Yeyati, Phys. Rev B \textbf{68}, 035105 (2003).

\bibitem{choi00}
Mahn-Soo Choi, C. Bruder and D. Loss, Phys. Rev. B {\bf 62}, 13569 (2000).

\bibitem{pala07}
M.G.~Pala, M.~Governale, and J.~K\"onig,
New J. Phys. \textbf{9}, 278 (2007).

\bibitem{noneqth}A. F. Volkov, Phys. Rev. Lett. \textbf{74}, 4730 (1995);
F.K.~Wilhelm, G.~Sch\"on, and A.D.~Zaikin, Phys.~Rev.~Lett.~{\bf 81}, 1682 
(1998);
S.-K. Yip, Phys. Rev. B {\bf 58}, 5803 (1998); 
P.~Samuelsson, J. Lantz, V. S. Shumeiko, and G. Wendin, Phys. Rev. B {\bf 62}, 
1319 (2000);  
E. V. Bezuglyi, V. S. Shumeiko, and G. Wendin, Phys. Rev. B {\bf 68}, 134506 
(2003);
F.~Giazotto, T.T.~Heikkil\"a, F.~Taddei, R.~Fazio, J.P.~Pekola, and 
F.~Beltram, Phys.~Rev.~Lett.~{\bf 92}, 137001 (2004).

\bibitem{baselmans99}
J.J.A.~Baselmans, A.F.~Morpurgo, B.J.~van Wees, T.M.~Klapwijk, 
Nature {\bf 397}, 43 (1999).

\bibitem{Bardeen}
J. Bardeen, Phys. Rev. Lett. {\bf 9}, 147 (1962);
B.D. Josephson, Phys. Lett. {\bf 1}, 251 (1962).

\bibitem{meir-wingreen}
Y. Meir and N.S. Wingreen, Phys. Rev. Lett. {\bf 68}, 2512 (1992).

\bibitem{koenig96}
J.~K\"onig, H. Schoeller, and G. Sch\"on, Phys.~Rev.~Lett. {\bf 76}, 1715 (1996);
J.~K\"onig, J. Schmid, H. Schoeller, and G. Sch\"on, Phys.~Rev.~B {\bf 54} 16820 (1996).

\bibitem{normalline} 
In the case of a normal line, the factor $|\Delta/\omega|$ is not present, but 
the argument remains the same.

\end{thebibliography}
\end{document}